\documentclass[journal]{IEEEtran}
\usepackage{graphicx} % Required for inserting images
\usepackage{amsmath}
\usepackage{amssymb}
\usepackage{multirow} 
\usepackage{booktabs}
\usepackage{url}
\usepackage{hyperref}
\usepackage{makecell}
\usepackage{float}
\usepackage{caption}
\usepackage[switch]{lineno}
\usepackage{color}
\usepackage{threeparttable}
\newcommand{\Rhat}{\ensuremath{\hat{R}}}

\title{
    Radiological mapping and uncertainty quantification by a fast Microcanonical Langevin Monte Carlo sampler
}
\author{
    Lei Pan, Jaewon Lee, Brian J.~Quiter, Jakob Robnik, Uro{\v{s}} Seljak, Jayson R.~Vavrek
    \thanks{
        L.~Pan, J.~Lee, B.J.~Quiter, and J.R.~Vavrek are with the Nuclear Science Division, Lawrence Berkeley National Laboratory, Berkeley, CA, 94720 USA.
        J.~Robnik and U.~Seljak are with the Physics Department, University of California at Berkeley, Berkeley, CA 94720, USA.

        This material is based upon work supported by the DOE Office of Science under contract DE-SC0024232, as well as by the U.S.\ Department of Energy, National Nuclear Security Administration, Office of Defense Nuclear Nonproliferation Research and Development (DNN R\&D).%
    }
}
\date{\today}

\begin{document}
%\linenumbers

\maketitle

\begin{abstract}
Radiological mapping plays a critical role in nuclear emergency response and environmental management activities.
A radiation image, representing the spatial and intensity distribution of the radioactivity, is reconstructed from the radiation data and the associated contextual information.
Typical image reconstruction methods, such as Maximum Likelihood Expectation-Maximization (ML-EM), only provide point estimates of the pixel or voxel activities without associated uncertainties.
Here, we apply a new Microcanonical Langevin Monte Carlo (MCLMC) sampler for radiation image reconstruction and uncertainty quantification.
The MCLMC sampler properties are first tested with synthetic radiation images.
Methods to obtain the radiation distribution estimate and the associated uncertainty from the samples drawn by MCLMC are discussed.
Given sufficient measurement statistics, the radiation distribution estimate obtained from MCLMC results closely resembles the ground truth with less risk of over- or under-fitting compared to ML-EM.
When MCLMC is run in parallel on a GPU, the samples can converge to the posterior distribution in about 10 seconds for an image with $10^3$--$10^4$ pixels, which is significantly faster than other comparable Markov Chain Monte Carlo (MCMC) samplers.
We also tested MCLMC on a dataset from a real distributed source radiological mapping campaign, and the reconstructed results agree well with the expected activity map.
The fast MCLMC sampler therefore enables improved imaging accuracy and prompt uncertainty quantification for reconstructed radiation images, which can better inform decision-making in response to radiological events.
\end{abstract}

\begin{IEEEkeywords}
    Microcanonical Langevin Monte Carlo, Markov Chain Monte Carlo, uncertainty quantification, radiation mapping, distributed sources
\end{IEEEkeywords}

\section{Introduction} 

Radiological mapping seeks to determine the spatial and intensity distribution of radioactive materials in an area under investigation, which can be an important informative step in various activities of nuclear security and consequence management.
In a mapping analysis, radiation detectors move throughout the environment to measure the counts of gamma rays or neutrons emitted from radioactive materials.
The radiation detector data is then associated with contextual information, measured by cameras or LiDAR instruments, to reconstruct a radiation map representing the spatial and intensity distribution of the radioactivity within the contextual model.
Such a process is known as Scene Data Fusion (SDF)~\cite{vetter2018gamma, vetter2019advances, hellfeld2021free, vavrek2024surrogateIII}.
In the image reconstruction process, the environment, i.e., the imaging domain, can be modeled as a partially-occupied 3D grid of voxels.
To achieve reasonable spatial fidelity, the voxel size is typically chosen to be relatively small compared to the scale of the scene, e.g., a voxel size of $10$--$200$~cm vs.\ a scene scale of tens or hundreds of meters.
This often results in many more voxels than measurements, meaning that the reconstruction is a high-dimensional and often highly underdetermined inverse problem.
Standard image reconstruction methods exist, such as Maximum Likelihood Expectation-Maximization (ML-EM) or Maximum a Posteriori Expectation-Maximization (MAP-EM), but only provide point estimates of the voxel activities without associated uncertainties~\cite{dempster1977maximum, hellfeld2019gamma}.
However, uncertainties of the reconstructed radioactivity can provide valuable information for interpretation of the reconstruction results, which can be pivotal for decision-making in many radiological events.

There has been exploration of methods for uncertainty quantification (UQ) of image reconstruction.
The challenge for most existing methods is the impractically high computation cost stemming from the high dimension (i.e., the number of pixels or voxels) of the problem.
For instance, Iterative Bayesian Unfolding performs analytical error propagation of the ML-EM algorithm, but the computation cost is prohibitively high, scaling as at least $O(I^2 J^3)$, where $I$ is the number of measurements and $J$ is the number of pixels or voxels~\cite{bourbeau2018pyunfold}.
From a Bayesian perspective, image reconstruction can be considered a \textit{maximum a posteriori} (MAP) problem and UQ can be naturally performed given the full posterior distribution~\cite{calvetti2018inverse}.
The posterior is often analytically intractable and different methods have been used to obtain the posterior. 
Markov Chain Monte Carlo (MCMC) sampling can be used to obtain the posterior~\cite{bardsley2012mcmc}, and this method has been applied to image reconstruction and UQ for medical imaging \cite{zhou2020bayesian, higdon1999bayesian} and radiological source detection and characterization~\cite{vavrek2021mcmc, lee2024radiation, dione2024utilizing}. 
{\color{black}MCMC methods  generate samples from the posterior distribution, and the resulting sample set can be used to approximate the posterior. The posterior depends on the forward physical model and the selected prior distribution. The MCMC estimated posterior should not be interpreted as an exact or model-independent posterior.}

However, existing MCMC samplers typically have a large computation time for high dimensional problems and near-real-time operation is challenging or impossible.
The Laplace approximation has also been used to estimate the posterior for UQ of radiation maps~\cite{rue2009approximate,lee2024radiation}.
Yet, compared to MCMC sampling, the Laplace approximation can be inaccurate when the posterior is strongly non-Gaussian near the mode (e.g., multimodal or skewed).
Methods that enable real-time or near-real-time UQ with good accuracy for posteriors with complex shapes are urgently needed.

In this work, we adopted a recently developed MCMC sampler, Microcanonical Langevin Monte Carlo (MCLMC)~\cite{pmlr-v253-robnik24a}, for radiological image reconstruction and UQ.
The MCLMC sampler is implemented in the Blackjax library~\cite{blackjax_mclmc} and we tested it with both synthetic and real radiological mapping data.
MCLMC uses physics-inspired methods for efficiently sampling high-dimensional distributions, making it attractive for UQ of high-dimensional reconstructed radiation images.
MCLMC, like other MCMC-based methods, also provides more convenient convergence properties.
For instance, traditional ML-EM has no well-defined stopping criterion, and the reconstruction could be under- or overfit to Poisson noise in the data depending on the number of iterations chosen.
{\color{black}In contrast, given sufficient data and adequate convergence, the reconstruction from MCLMC closely resembles the true image without concern of choosing a stopping criterion, and hence has less risk of under- or overfitting. Despite this advantage, the resulting posterior remains conditional on the selected prior.
As with all Bayesian methods, and with MAP-EM, there is some arbitrariness when choosing the prior, but physics-informed priors that incorporate practical physical information (e.g., non-negativity constraints) can be used.}
Here, we show that a simple independent Gaussian prior performs well in terms of reconstructed image quality metrics, and that a Gaussian Process Prior (GPP)~\cite{williams2006gaussian}---which takes into account spatial correlations between pixels---outperforms the independent Gaussian prior.
We also show that MCLMC significantly outperforms other MCMC samplers available in the same Blackjax toolkit in terms of computational cost.
For instance, for a scenario with $J=480$ voxels and equal numbers of effective samples representing good convergence, MCLMC has a computation time of $13$~s while the prior state-of-the-art Hamiltonian Monte Carlo (HMC) needs $474$~s.
The sampling time of MCLMC can be further reduced by running in parallel on a GPU, where, for example, it converges in ${\sim}11$~s even at $J \sim 10^4$ voxels.
The reconstruction of a dataset from real radiological mapping by MCLMC agrees well with the expected activity map, and also provides uncertainty maps.
In conclusion, MCLMC is capable of near-real-time image reconstruction and UQ, which can enable improved decision-making based on radiological mapping.

\section{Methods}

\subsection{Forward model}
In a radiological mapping analysis, a radiation detector moves through an area to be mapped and conducts a series of~$I$ radiation count measurements in order to reconstruct the emission rates in the image or mapping space, which we discretize into~$J$ discrete pixels or voxels.
The count data are expressed as a data vector $\boldsymbol{x}^{[I \times 1]} = [x_{1}, x_{2}, \ldots , x_I ]^{T}$, where $x_i$ is the number of counts (typically gamma-ray or neutron) within the $i^\text{th}$ measurement dwell time $t_i$.
The counts are a Poisson sample
\begin{align}
    \boldsymbol{x} \sim \text{Poisson}(\boldsymbol{\lambda})
\end{align}
of the mean count vector $\boldsymbol{\lambda}^{[I \times 1]}$ (each element noted as ${\lambda}_i$), where
\begin{align}
    \boldsymbol{\lambda} = \boldsymbol{V} \boldsymbol{w}+b \boldsymbol{t}
    \label{eq:lambda} 
\end{align}
is the ``forward projection'' of the voxel intensities $\boldsymbol{w}$ to be reconstructed, and a background rate~$b$, which we assume to be constant and either known or reconstructed along with the~$\boldsymbol{w}$.
The negative log likelihood or ``Poisson loss'' $\ell(\boldsymbol{x} | \boldsymbol{\lambda})$ for obtaining a particular measurement~$\boldsymbol{x}$ given the mean counts~$\boldsymbol{\lambda}$ is
\begin{align}
\ell(\boldsymbol{x}\mid\boldsymbol{\lambda})
&= \sum_{i=1}^{I}\Big( \lambda_i - x_i \log \lambda_i + \log\Gamma(x_i+1)\Big)
\label{eq:log_likelihood}
\end{align}
where $\Gamma(\cdot)$ is the gamma function.
The system matrix $\boldsymbol{V}^{[I \times J]}$ (with units of time) incorporates the geometric and detector efficiency of the~$I$ measurements relative to each of the~$J$ image voxels, and its $(i, j)$ element is given by
\begin{align}
    V_{ij} \equiv \frac{\eta_{ij} t_i}{4 \pi r_{ij}^2},
\end{align}
where $\eta_{ij}$ is the effective area of the detector to voxel~$j$ at timestamp~$i$, and $r_{ij}$ is the corresponding scalar distance.
We assume here for simplicity that there is no attenuation by intervening materials between the radioactive source and detector, although the incorporation of attenuation by introducing an $\exp \left(- \int_{\vec{r}_i}^{\vec{r}_j} \mu(\vec{r}) d \vec{r} \right)$ term in the numerator is possible if the linear attenuation coefficient map $\mu(\vec{r})$ is known.
$\boldsymbol{V}$ is a known parameter as it can be calculated from the measurement setup and detector effective area.
The measurement time durations at each detector position $\boldsymbol{t}^{[I \times 1]}$ are known from the measurement process, and the total measurement time is $T = \sum_{i} t_i$.
It is worth noting that the number of measurements~$I$ is independent of the number of reconstructed image voxels~$J$, and thus that $\boldsymbol{V}$ is not in general square.
Additionally, in practice, the effective area $\eta_{ij}$ is a function of photon energy $E$, and thus the counts $\boldsymbol{x}$ are defined as the counts within some energy region of interest around a given energy~$E$.

\subsection{Image reconstruction}\label{sec:image_reconstruction}
In traditional ML-EM reconstruction, point estimates of intensities~$\boldsymbol{w}$ and background rate~$b$ are found by an iterative minimization of the negative log likelihood in Eq.~\ref{eq:log_likelihood}~\cite{dempster1977maximum}.
The ML-EM iterative process is guaranteed to monotonically reduce the Poisson loss.
Conversely, in a Bayesian radiological image reconstruction framework, we can write Bayes' theorem as
\begin{align} 
p(\boldsymbol{w}, b \,|\,  \boldsymbol{x}) \propto p(\boldsymbol{x} | \boldsymbol{\lambda}) \cdot p(\boldsymbol{w}, b) 
    \label{eq:bayes}
\end{align}
where $p(\boldsymbol{w},b \,|\, \boldsymbol{x})$ is the joint posterior distribution of~$\boldsymbol{w}$ and~$b$, $p(\boldsymbol{x} | \boldsymbol{\lambda})$ is the likelihood of observing the data, and $p(\boldsymbol{w}, b)$ is the joint prior distribution of the (non-negative) intensities~$\boldsymbol{w}$ and background rate~$b$.
Various priors may be chosen.
{\color{black}One natural choice of prior is that~$\boldsymbol{w}$ and~$b$ follow Gaussian distributions (truncated at~$0$), and that the elements of~$\boldsymbol{w}$, representing individual voxels, are independent of one another, which we call the truncated Gaussian prior. The truncated Gaussian prior assumes that individual pixels of the radiation field are independent and non-negative, following Gaussian distributions. }
{\color{black}The means for~$\boldsymbol{w}$ may be chosen as their respective point estimates from an initial ML-EM reconstruction, and the standard deviations for $\boldsymbol{w}$ may be set equal to those means or multiples thereof. The background rate~$b$ is treated as a well-known prior as it can be measured or reconstructed by ML-EM, so the mean of~$b$ can be set to the measured or ML-EM reconstructed value with a standard deviation significantly smaller than the mean (e.g., standard deviation set as mean/100).}
{\color{black}An alternative prior we consider is a Gaussian Process Prior (GPP), which, as demonstrated in previous work~\cite{lee2024radiation}, can significantly improve image reconstruction over ML-EM due to its inclusion of spatial correlations among voxel values. The GPP assumes nearby pixels have stronger correlation than distant pixels in the radiation field, and that all pixels have non-negative intensity.} {\color{black}The GPP implemented in this work does not include the background rate term~$b$, but it will be shown that the GPP prior can still provide good estimates of the intensity map when~$b$ is negligible compared to the source intensity.}
{\color{black}Different priors can represent different physical features of the radiation field and the prior should be chosen accordingly.}

To obtain the full posterior distribution for UQ, an MCMC sampler generates samples that converge to the posterior given sufficient number of samples.
There are different MCMC samplers available in the Blackjax library~\cite{blackjax_mclmc}, such as the standard HMC, the No U-Turn Sampler (NUTS), which is a state-of-the-art variant of HMC, and so on.
A well-designed MCMC sampler should converge robustly and quickly at high dimensions.
The recently developed MCLMC, also available in Blackjax, has been demonstrated to significantly outperform HMC and NUTS~\cite{pmlr-v253-robnik24a}---it is orders of magnitude faster in high dimensions than HMC and NUTS, and features automatic hyperparameter tuning and extremely fast burn-in.
These properties of MCLMC are expected to be well-suited for the challenging problem of radiological image reconstruction and UQ because of the potentially large number of dimensions (pixels or voxels)~$J$, which is often in the tens or hundreds of thousands and could exceed~$10^7$ in particularly large problems.

The MCLMC sampler draws latent samples $\boldsymbol{\phi}$ in a $(J+1)$-dimensional phase space, where each element of the sample is a real number:
\begin{align}
\boldsymbol{\phi} = 
\begin{bmatrix}
\boldsymbol{\phi}_w \\
\phi_b
\end{bmatrix}
\in \mathbb{R}^{J+1}.
\label{eq:phi_definition}
\end{align}
$\boldsymbol{\phi}_w$ is a vector with $J$ elements noted as $\boldsymbol{\phi}_{w,j}$.
Because the intensities $\boldsymbol{w}$ and background rate $b$ are non-negative, we use a link function to apply the non-negativity constraint.
For example, one choice of the link function takes the element-wise exponential value of the latent sample $\boldsymbol{\phi}$ as the value of $\boldsymbol{w}$ and $b$.
Mathematically, we can write the log posterior $\log p(\boldsymbol{\phi} \,|\, \boldsymbol{x})$ as 
\begin{align}
\log p(\boldsymbol{\phi} \mid \boldsymbol{x})
\propto\ 
&\log p\!\left(\boldsymbol{x} \mid \boldsymbol{\lambda}= f\!\left(\exp(\boldsymbol{\phi}_w),\, \exp(\phi_b)\right)\right) \nonumber \\
&+ \log p\!\left(\exp(\boldsymbol{\phi}_w),\, \exp(\phi_b)\right)
+ \sum_{j=1}^{J}\phi_{w,j} + \phi_b
\label{eq:link_function}
\end{align}
where $\sum_{j=1}^{J}\phi_{w,j} + \phi_b$ is the Jacobian determinant corresponding to the link function based on the change of variable principle.
After drawing the latent samples $\boldsymbol{\phi}$, we apply the link function to map them back to the non-negative problem variables.

To understand the convergence performance of the MCMC samplers, we use the~$\Rhat$ statistic (also known as the Gelman-Rubin statistic) and the Effective Sample Size (ESS)~\cite{roy2020convergence}.
The $\Rhat$ is defined for multiple independent chains started from dispersed initial values, and compares the between-chain variance and the within-chain variance.
The $\Rhat$ is larger than~$1$ by definition.
An $\Rhat < 1.1$ is generally considered good convergence, and smaller $\Rhat$ is recommended for higher-precision results.
Because samples produced by MCMC are auto-correlated, the number of effective independent samples is smaller than the total number of samples.
The ESS estimates the number of effective independent samples, and an ESS larger than 200 usually indicates good convergence~\cite{roy2020convergence}.
We also use the trace plot and autocorrelation plot for visual inspection of the convergence behavior.
The trace plot displays the sampled parameter values against the sample index, and a well-converged chain should show noisy but stationary behavior without long term trends.
The autocorrelation plot displays the autocorrelation function (ACF) of a sampled parameter as a function of lag, where the lag refers to the number of sample indices by which the data is shifted when computing the correlation between an MCMC chain and a delayed version of itself~\cite{roy2020convergence}.
For a well-converged chain, the ACF should drop to 0 quickly (e.g., within a few tens of lags) and then oscillate around 0.
ACF staying high and positive for a long time indicates strong correlation and poor convergence.

With the samples drawn by an MCMC sampler, we have a $(J+1)$-dimensional joint posterior distribution that cannot immediately be viewed as a simple 2D map of reconstructed intensities or uncertainties---some data reduction method must be applied.
In some cases it is convenient to use a single representative value of the marginal distribution as the point estimate of intensity for each pixel.
For a symmetric marginal distribution where the mean, median and mode are the same, such a representative value can be naturally set as the mean.
However, for pixels with highly skewed marginal distributions, there may be a large difference between the mean, median, and mode (as seen in Fig.~\ref{fig:mclmc prior and post}), and thus different reductions may be used for point estimates of the intensity.
Here, we compare three methods of reducing the posterior as the point intensity estimate: 1) using the marginal mean of each pixel, 2) using the marginal mode of each pixel, 3) using the individual components of the multi-dimensional joint mode.

Theoretically, the most likely set of intensity estimates is the joint mode in the multi-dimensional space, not the mean or mode of the marginal distribution.
We can use the sample with the highest log posterior probability as an approximation of the joint mode, and the components of this approximated joint mode are used as the point intensity estimate for each pixel.
This approximated joint mode drawn by the MCMC sampler is not guaranteed to coincide with the theoretical joint mode, as MCMC generates samples according to the posterior distribution rather than performing optimization.
The theoretical joint mode can be obtained by directly maximizing the posterior; here we use the L-BFGS optimizer~\cite{liu1989limited}.
Then, each component of the theoretical joint mode may be used as the point intensity estimate for each pixel.
We note that this optimization approach on its own only provides a point estimate without uncertainty of the estimated intensity, though uncertainty information is available from the MCMC sampler.

Although we are interested in reductions that may be used as the point estimate of the intensity and uncertainty for each pixel, we emphasize again that the posterior distribution is multi-dimensional and that the 2D intensity and uncertainty maps are just condensed representations of the posterior.
To characterize and understand the performance of MCLMC in the following analyses, we choose to use the marginal mean as the point intensity to conveniently visualize the posterior when different parameters of MCLMC are used.
For uncertainty estimate, since the marginal distribution for some pixels can be highly skewed, we use the $68\%$ highest density interval (HDI), the narrowest interval that contains the most probable $68\%$ of the posterior values, of the marginal distribution, as the point uncertainty estimate of each pixel.
This is analogous to the $\pm 1$ standard deviation credible interval for a Gaussian distribution.
{\color{black} We note that the uncertainty estimate discussed in this work is conditional on a selected prior and it does not include uncertainty in prior selection and prior hyperparameter estimation. }

To compare the similarity between the reconstructed 2D intensity map and the ground truth, we used the Peak Signal-to-Noise Ratio (PSNR) and the Structural Similarity Index Measure (SSIM) as quantitative metrics.
A larger PSNR means better similarity~\cite{hore2010image}.
The SSIM used here is the standard compact form that implicitly combines luminance, contrast, and structure~\cite{wang2004image}.
Then a mean SSIM is obtained as the spatial average of the pixel-wise SSIM.
The mean SSIM has values from $-1$ to $1$ by definition, and identical images have an SSIM of~$1$.

\subsection{Synthetic radiological mapping data}

\begin{figure}[!htbp]
    \centering
    \includegraphics[width=1\linewidth]{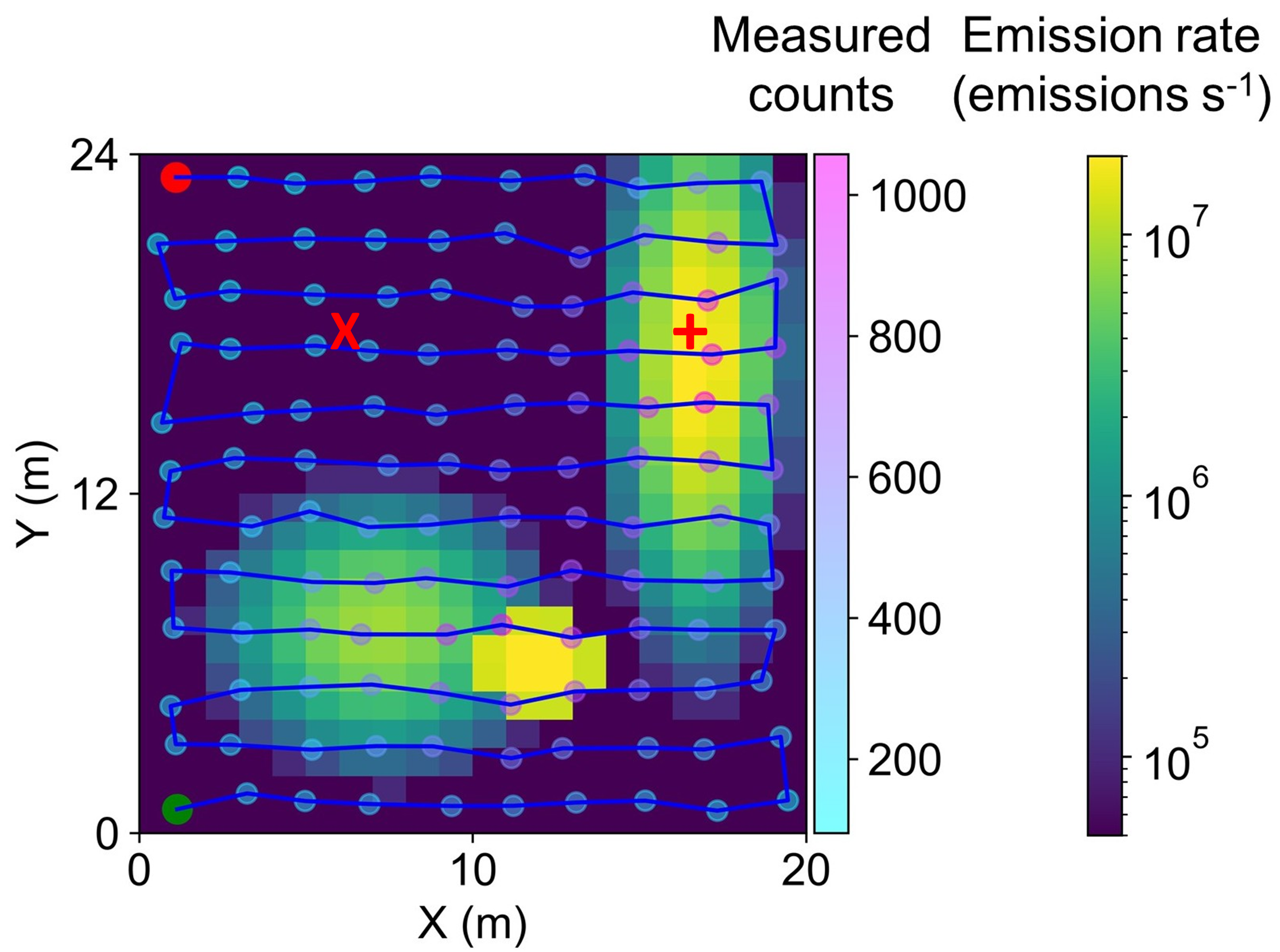}
    \caption{
        Synthetic radiological mapping scenario in which a detector rasters over a field containing a distributed radiological source consisting of three plumes.
        Each dot represents a $t_i=5$~s duration measurement position and is colorized by the measured counts in that time interval.
        The pixel sizes are set to 1~m$^2$.
        The detector trajectory starts from the green dot near $(0, 0)$ and stops at the red dot near $(0, 24)$.
        The $\mathsf{X}$ and $+$ represent a typical pixel in the background region and in the source region, respectively.
    }
    \label{fig:synthetic_data_ground_truth}
\end{figure}

To test MCLMC, we generated synthetic 2D radiological mapping data, as shown in Fig.~\ref{fig:synthetic_data_ground_truth}.
The synthetic ground truth intensities~$\boldsymbol{w}$ (with $J=480$ pixels) consist of three Gaussian-plume-like distributions of different shapes, intensities, and separations, which are confined to a flat plane at a height of $z=0$~m.
A constant small but non-zero background rate of $b = 0.1$~counts/s was modeled, which provides a high but non-infinite signal to background ratio.
An isotropic radiation detector with diameter $1$~cm and $100\%$ intrinsic efficiency (i.e., $\eta \equiv 0.785$~cm$^2$) moves in a raster pattern with line spacing of $2$~m at a constant $3$~m above the radioactive source plane.
The raster lines are noised with a standard deviation of $0.2$~m in all three dimensions to mimic real position fluctuations and avoid potential aliasing effects.
In reality, the detector moves continuously along its trajectory, but computationally, the trajectory is discretized into~$I$ distinct measurement positions with dwell times~$t_i$.
To generate measurement data, we assume that, at each measurement position, the detector remains fixed for a dwell time~$t_i$ to measure the counts due to the source and background terms.
Using Eq.~\ref{eq:lambda}, the mean counts $\boldsymbol{\lambda}$ at each position is calculated, and the measured counts at each position~$\boldsymbol{x}$ are then drawn from a Poisson distribution with mean $\boldsymbol{\lambda}$.
For a field with the dimensions given in Fig.~\ref{fig:synthetic_data_ground_truth}, in a typical scenario, we set the total measurement time $T = \sum_i t_i = 10$~minutes, such as $120$ measurement positions discretized to $t_i = 5$~s.
In the following sections, however, we will also examine algorithm performance at measurement times ranging from $T=1$~minute to $T=1000$~minutes.
At this scale, $T = 1$~minute is somewhat short, likely requiring vehicle-based operation to attain movement speeds on the order of $4$~m/s.
$T=1000$ minutes is unrealistically long due to operational considerations such as system battery life, but serves as a useful ideal case.
Between those limits, the duration of $T=100$~minutes may also be impractical, but can be useful as an equivalent case to $T=10$~minutes with a $10\times$ more efficient detector.

\section{Results}
\subsection{MLEM reconstruction}

\begin{figure*}[!htbp]
    \centering
    \includegraphics[width=1\linewidth]{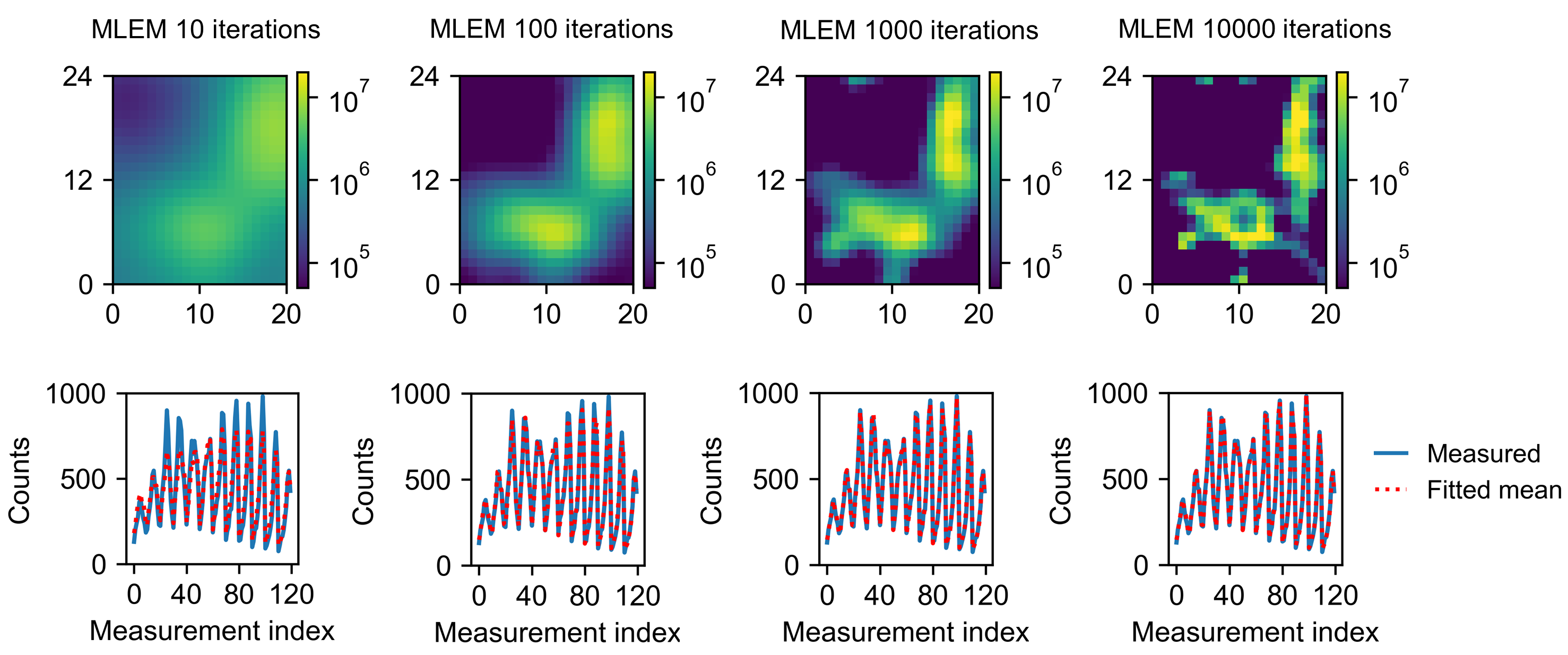}
    \caption{
        Reconstruction by ML-EM with different number of iterations.
        Total measurement time $T=10$~minutes.
        Top row: reconstructed intensity distributions.
        Bottom row: measured counts vs.\ mean counts calculated from Eq.~\ref{eq:lambda} using the reconstructed intensities.
        PSNR/SSIM is 11.79/0.23, 16.61/0.59, 23.50/0.89, 19.21/0.64, respectively, for 10, 100, 1000, 10000 iterations.
    }
    \label{fig:mlem}
\end{figure*}

\begin{figure}[!htbp]
    \centering
    \includegraphics[width=1.0\linewidth]{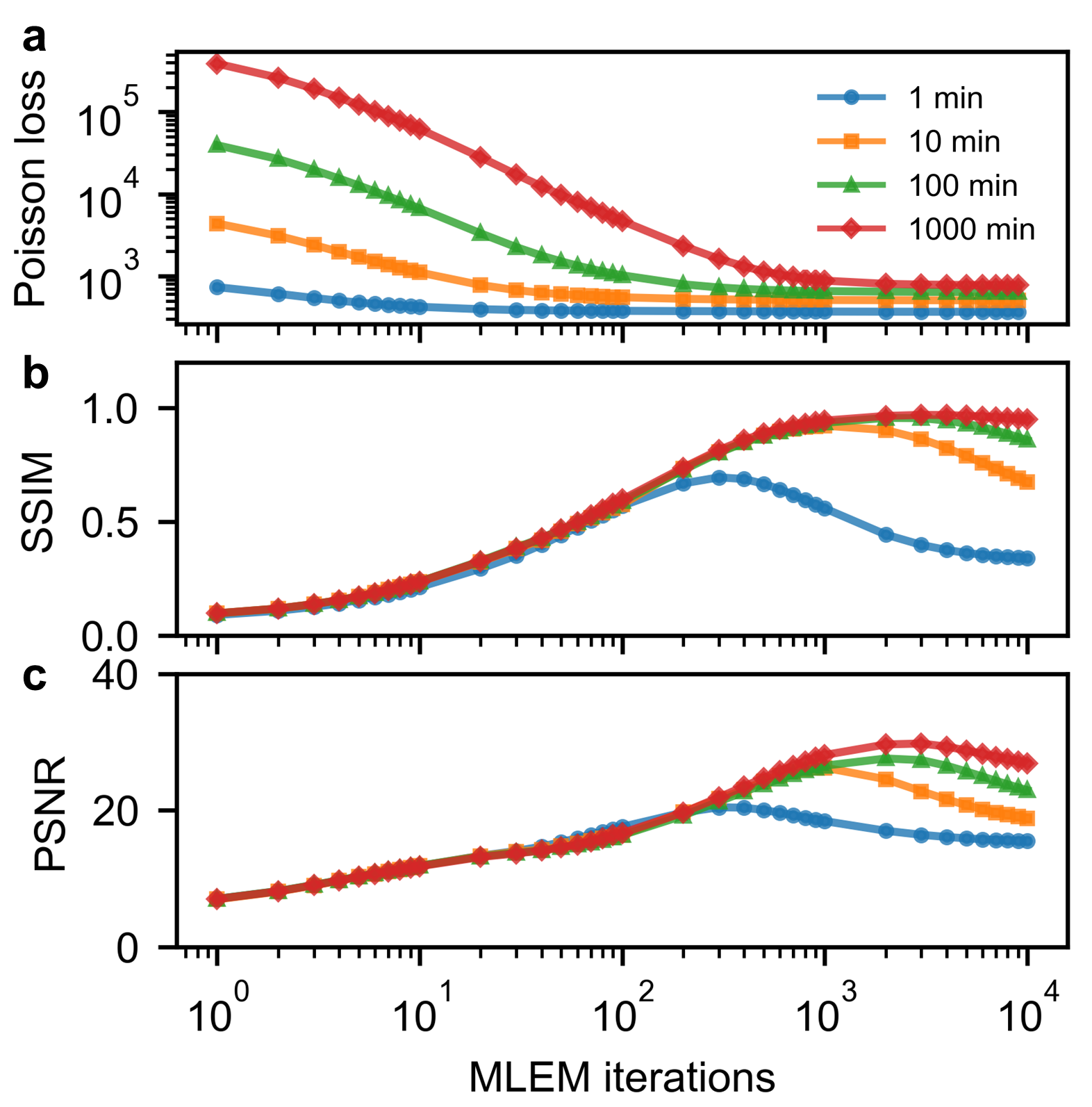}
    \caption{
        Performance metrics vs.\ number of ML-EM iterations for different total measurement times~$T$.
        (a) Poisson loss calculated using the ML-EM reconstructed intensities.
        (b) SSIM and (c) PSNR from comparison of reconstructed image at different number of ML-EM iterations against the ground truth.
        The overall highest PSNR = 29.79 and SSIM = 0.97 are obtained at 3000 iterations for $T=1000$~minutes.
    }
    \label{fig:psnr_ssim_vs_iterations}
\end{figure}

We first demonstrate traditional ML-EM reconstruction of the synthetic data and then compare it with MCLMC.
We initialize ML-EM with a flat image, i.e., the same initial value of $1\times 10^7$ emissions/s for all pixels.
Fig.~\ref{fig:mlem} shows the ML-EM-reconstructed intensities with different numbers of iterations for a total measurement time of $T=10$~minutes, i.e., $120$ measurements of $t_i=5$~s.
Using the reconstructed intensities, the fitted mean counts calculated with Eq.~\ref{eq:lambda} are compared with the measured counts (see the second row of Fig.~\ref{fig:mlem}).
With 10 ML-EM iterations, the mean counts have a large mismatch to the measured counts, but as the number of ML-EM iterations increases, by the nature of the ML-EM algorithm, the fitted mean counts show better agreement with the measured counts.
Quantitatively, the Poisson loss calculated from Eq.~\ref{eq:log_likelihood} decreases as a function of the number of ML-EM iterations (see Fig.~\ref{fig:psnr_ssim_vs_iterations}a).
However, we cannot tell which number of iterations is optimal from comparisons in count space alone due to potential overfitting.

Fig.~\ref{fig:psnr_ssim_vs_iterations} shows the PSNR and SSIM of the ML-EM reconstructed intensities against ground truth for different ML-EM iterations with different total measurement time.
Both the PSNR and SSIM first increase to a maximum and then decrease, which means the ML-EM reconstruction is first underfit and then overfit.
Increasing the total measurement time~$T$ directly improves measurement statistics and thus the reconstruction performance.
The optimal PSNR and SSIM are improved and both metrics stay near their maxima for a wider range of ML-EM iterations.
However, in real radiological mapping scenarios, it is not possible to select the optimal number of iterations for ML-EM based on the ground truth, since it is not known.

\subsection{MCLMC reconstruction and uncertainty quantification}\label{sec:results_recon}

\begin{figure}[!htbp]
    \centering
    \includegraphics[width=1\linewidth]{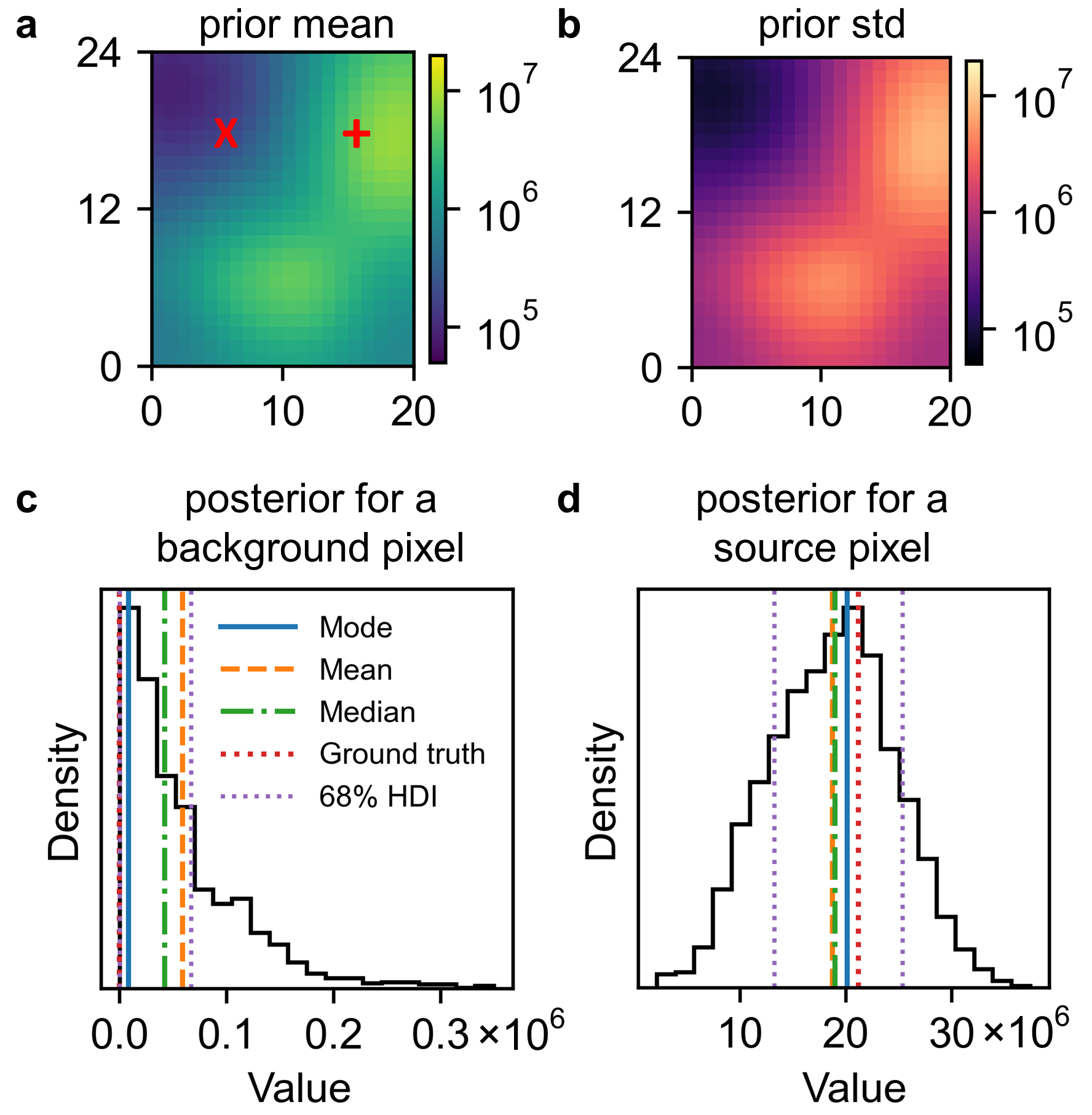}
    \caption{
        Prior and posterior distributions of MCLMC for $T=100$~minutes.
        (a),~(b) Maps showing the mean and standard deviation of the truncated Gaussian prior distribution for all pixels, from $10$ ML-EM iterations.
        The $\mathsf{X}$ and $+$ represent a typical pixel in the background region and in the source region, respectively (same location as in Fig.~\ref{fig:synthetic_data_ground_truth}).
        (c),~(d) Marginal posterior distributions for a pixel in background/radioactive source region shown in (a).
        HDI: Highest Density Interval.
    }
    \label{fig:mclmc prior and post}
\end{figure}

\begin{figure*}[!htbp]
    \centering
    \includegraphics[width=1\linewidth]{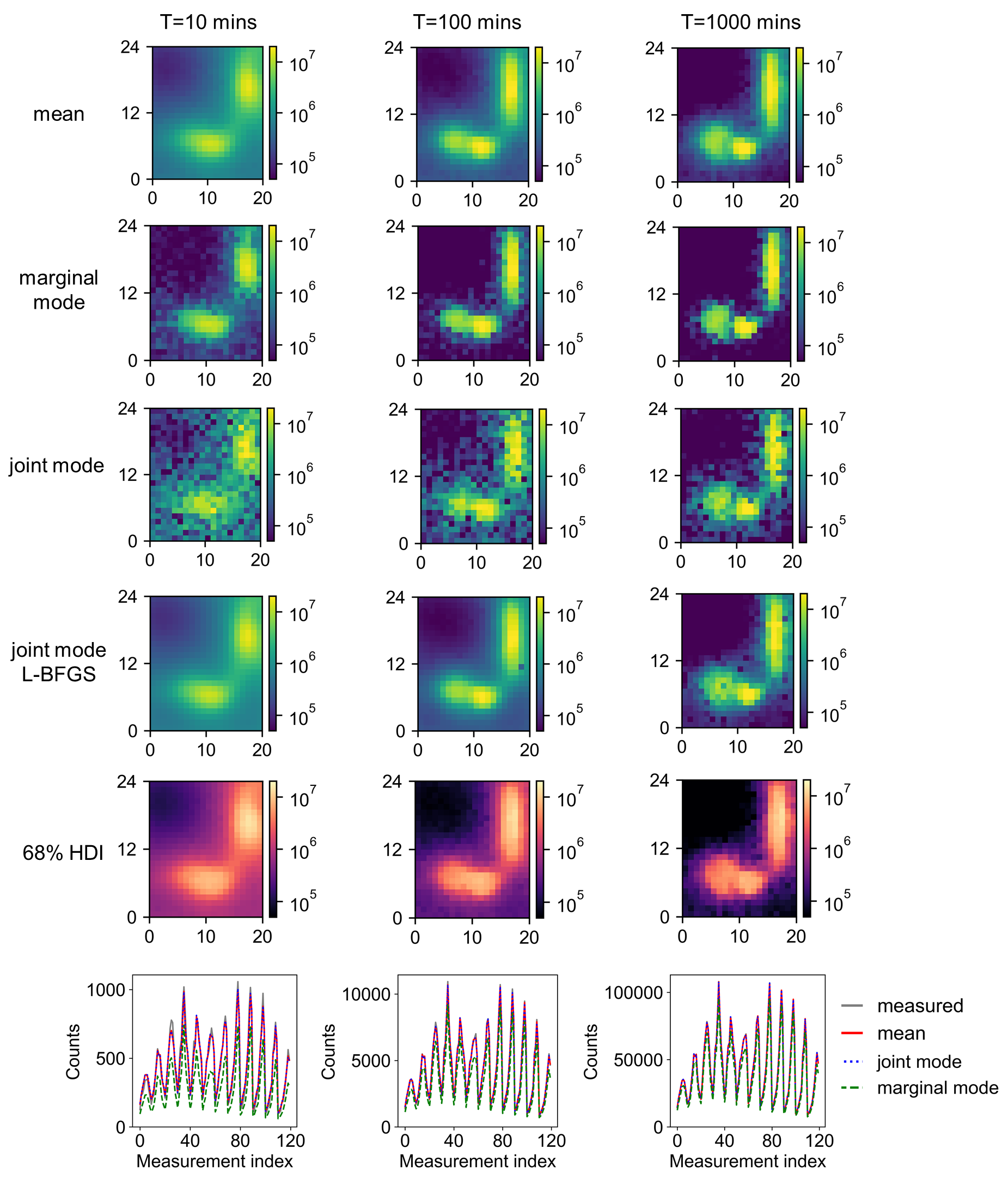}
    \caption{
        Intensity and uncertainty maps.
        Top 4 rows: intensity maps showing different properties of the posterior distribution.
        Row 5: $68\%$ Highest Density Interval of marginal distributions for individual pixels.
        Row 6: measured counts vs.\ mean counts calculated from Eq.~\ref{eq:lambda} using the value of mean, marginal mode and joint mode in the top 3 rows.
        Image quality metrics are given in Table~\ref{tab:mean_vs_mode_at_different_measurement_time}.
    }
    \label{fig:mclmc mean vs mode and fitting vs mea}
\end{figure*}

MCLMC was used to reconstruct the intensities~$\boldsymbol{w}$ from the synthetic data in Fig.~\ref{fig:synthetic_data_ground_truth}. 
We used the point estimates from ML-EM (10 iterations) as both the mean and standard deviation of the truncated Gaussian prior distribution---see Fig.~\ref{fig:mclmc prior and post}(a) and (b).
The estimates from ML-EM with too many iterations represent a possible image that overfits the measurement (such as 1000 iterations in Fig.~\ref{fig:mlem}).
Hence, a weakly-informative prior without overfitting, such as ML-EM with a low number of iterations, should be used to avoid having the MCLMC posterior get stuck around the overfit image.
We then ran~$10000$ MCLMC samples drawn from the $(J+1)$-dimensional phase space, where $J=480$.
{\color{black}We first demonstrate that the obtained posterior is appropriate through comparison of the posterior to ground truth. However, such a comparison is not straightforward because of the multi-dimensional posterior. We selected different representative quantities for the marginalized posterior (Fig.~\ref{fig:mclmc prior and post}) and multi-dimensional posterior (Fig.~\ref{fig:mclmc mean vs mode and fitting vs mea}) to show that the obtained posterior is appropriate.}
Fig.~\ref{fig:mclmc prior and post}(c) and (d) show the marginal distribution of two representative pixels, one pixel with low intensity in the background region (see the location of the $\mathsf{X}$ in Fig.~\ref{fig:synthetic_data_ground_truth} and in Fig.~\ref{fig:mclmc prior and post}(a)) and another with high intensity in the radioactive source region (see the location of the $+$ in Fig.~\ref{fig:synthetic_data_ground_truth} and in Fig.~\ref{fig:mclmc prior and post}(a)).
The histogram of sample values for the pixel inside the source region is approximately symmetric, and the mean, median and mode are similar. 
These metrics of the posterior distribution are close to the ground truth intensity for the selected source pixel---the mean is $1.87\times 10^7$ and the ground truth is $2.12\times 10^7$.
In contrast to the source pixel, the distribution of the sample values for the background pixel is highly skewed, which leads to significant differences among the mean, median and mode (mean = $5.9\times 10^4$, median = $4.2\times 10^4$, mode = $8.7\times 10^3$, ground truth = $0$).
This highly skewed distribution is due to the non-negativity constraint of pixel intensities and the true pixel intensity value of zero.

\begin{figure}[!htbp]
    \centering
    \includegraphics[width=1\linewidth]{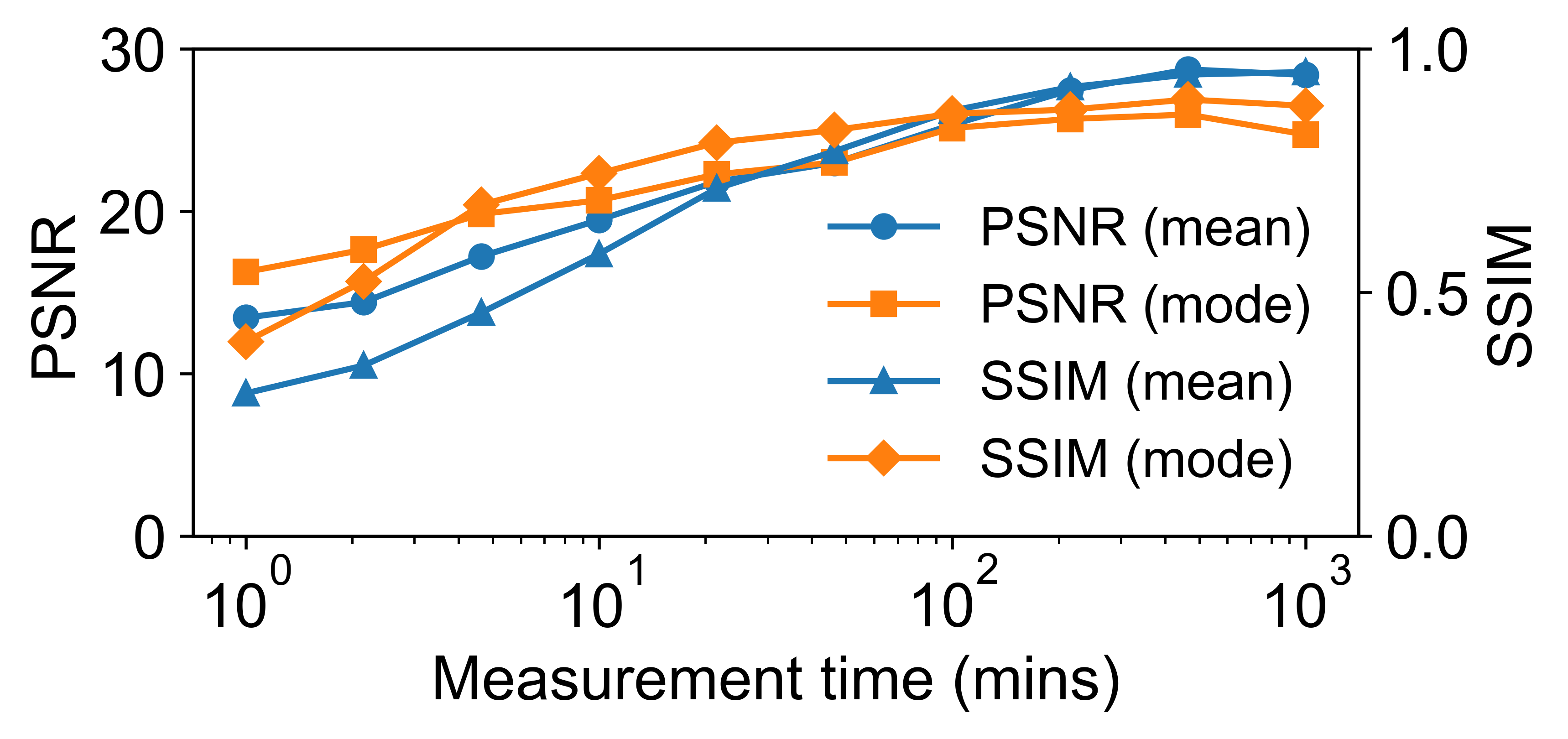}
    \caption{
        PSNR and SSIM calculated from the reconstructed intensity map using mean and marginal mode against the ground truth as a function of total measurement time~$T$.
    }
    \label{fig:MCLMC_mode psnr ssim vs measure time}
\end{figure}

Fig.~\ref{fig:mclmc mean vs mode and fitting vs mea} compares the point intensity maps using the marginal mean, marginal mode, joint mode approximated by the single MCLMC sample with the highest log posterior, and joint mode by optimization with the L-BFGS algorithm.
The marginal mode map has a lower intensity in the background region than the mean map, which makes it more similar to the ground truth than the mean map when the measurement time is relatively short, e.g., shorter than $T = 10$~minutes as shown in Fig.~\ref{fig:MCLMC_mode psnr ssim vs measure time}.
However, when the measurement time is relatively long, e.g., longer than $T=100$~minutes, the mean map is more similar to the ground truth than the mode (Fig.~\ref{fig:MCLMC_mode psnr ssim vs measure time}).

\begin{figure}[!htbp]
    \centering
    \includegraphics[width=1\linewidth]{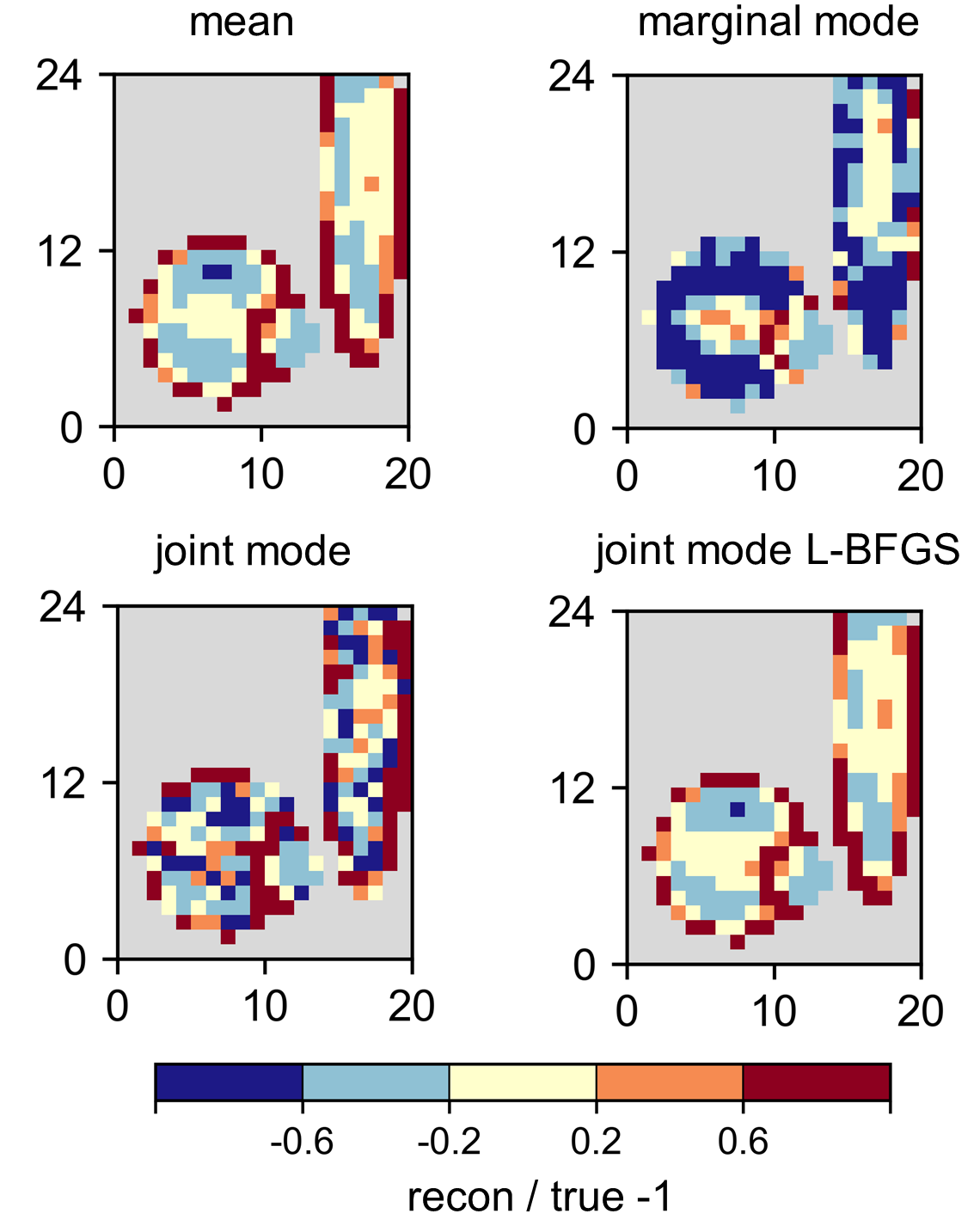}
    \caption{
        Relative difference for reconstructed intensity maps with total measurement time of $T=100$~mins against the ground truth.
        The pixels colored gray correspond to a true activity of $0$.
    }
    \label{fig:relative signed difference map T=100mins}
\end{figure}

The marginal mode is not the most likely intensity estimate.
As a result, the fitted mean counts calculated using the marginal mode are lower than the measured counts (see bottom row of Fig.~\ref{fig:mclmc mean vs mode and fitting vs mea}).
The marginal mode is usually not equal to the components of the joint mode for a multi-dimensional distribution, where the joint mode represents the most likely intensity estimate with highest log posterior probability.
The joint mode approximated by an MCLMC sample shows a pattern similar to the mean map and mode map, but it is noisy.
The fitted mean counts calculated using the approximated joint mode agree well with the measured counts.
The theoretical joint mode obtained by optimization with L-BFGS is very similar to the mean map, which means the joint mode coincides with the mean for this posterior.
Fig.~\ref{fig:relative signed difference map T=100mins} shows the relative difference between the intensity maps for a total measurement time of $T = 100$~minutes and the ground truth.
For the relative difference map of the mean, the center region of the source has smaller relative difference than the periphery region.
The relative difference map of the joint mode by L-BFGS is similar to that of the mean.
For the relative difference map of the marginal mode, the periphery region of the source area generally has smaller intensity than the ground truth.
The relative difference map of the joint mode is noisy since the joint mode map is noisy.

As the measurement statistics improve with longer measurement time, all intensity maps using different reductions, i.e., mean, marginal mode, joint mode, improve towards the ground truth.
As shown in Fig.~\ref{fig:MCLMC_mode psnr ssim vs measure time}, the PSNR and SSIM calculated from the intensity map using the mean and marginal mode against ground truth increase approximately asymptotically as the total measurement time increases.
The difference between the intensity maps also becomes smaller as the total measurement time increases, which is shown quantitatively in Table~\ref{tab:mean_vs_mode_at_different_measurement_time}.
In Table~\ref{tab:mean_vs_mode_at_different_measurement_time}, we also calculate the total intensity ratio, i.e., the ratio of reconstructed total intensity to the ground truth total intensity.
The reconstructed total intensities, using the point intensity estimate from the mean, the joint mode by MCLMC, the joint mode by L-BFGS, are larger than the ground truth total intensity.
The reconstructed total intensity using the marginal mode is smaller than the ground truth total intensity.
As the total measurement time increases, all total intensity ratios using different point intensity estimates improve towards~$1$.
Fig.~\ref{fig:mclmc mean vs mode and fitting vs mea} also shows the evolution of the uncertainty map, where the uncertainty for each pixel is the $68\%$ HDI.
It can be seen that the absolute uncertainty is larger for pixels with larger intensity.
With better measurement statistics, the uncertainty map becomes sharper, similar to the image itself, with smaller uncertainties in the background area and less blurring of source-area uncertainties.

\begin{figure}[!htbp]
    \centering
    \includegraphics[width=1\linewidth]{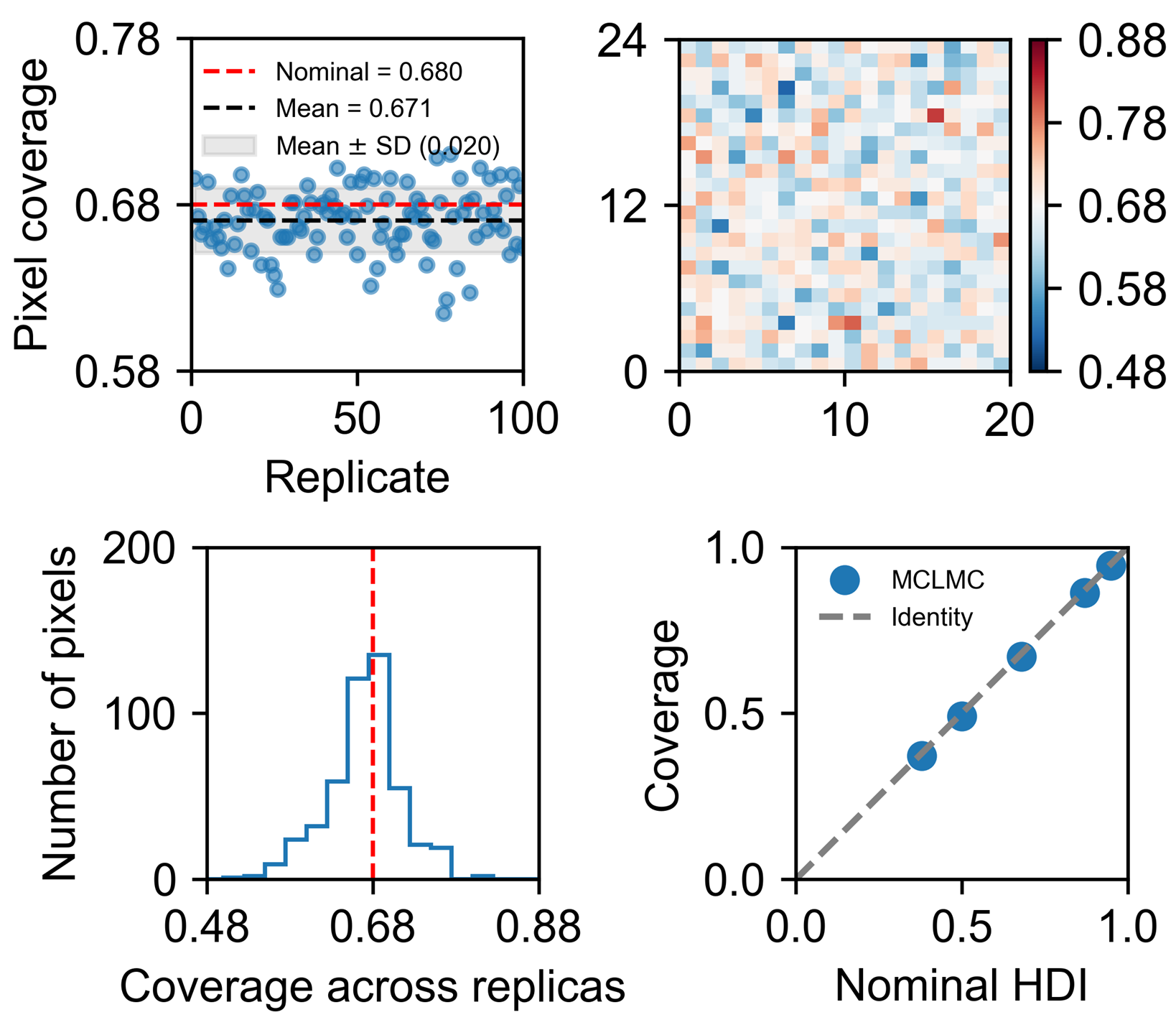}
    \caption{
       Posterior uncertainty calibration. Top left: pixel coverage for different replicas with nominal 68\% HDI (SD: standard deviation). Top right: the per-pixel coverage map across all replicas. Bottom left: the histogram for number of pixels with different coverage across replicas. Bottom right: the coverage vs.\ nominal HDI plot.
    }
    \label{fig:uncertainty calibration}
\end{figure}

{\color{black}We performed uncertainty calibration by conducting a full prior predictive experiment. We used the truncated Gaussian prior with a fixed mean and standard deviation of $10^{7}$~emissions/s for the individual pixels. Then ground truth pixel values were drawn from the prior distribution, and measured counts were generated using the process described in Fig.~\ref{fig:synthetic_data_ground_truth}. With the measurement data, posterior samples were drawn by MCLMC sampler with the same truncated Gaussian prior. A total of 100 replications were used, where the detector trajectory, system matrix and measurement time at each measurement position were fixed.  
We examined the coverage for all pixels across different replicas (see Fig.~\ref{fig:uncertainty calibration}). For a nominal 68\% HDI, the coverage per replicate is stable around 0.68, and the mean coverage for all replicas is 0.671 with a standard deviation of 0.020 that agrees with the nominal 0.68. The per-pixel coverage across different replicas shows that most pixels have a coverage of 0.68. When different nominal HDI values (HDI = 38\%, 50\%, 68\%, 87\%, 95\%) were used, the mean coverage for all replicas vs.\ the nominal HDI value is very close to the identity line. These results demonstrate that the posterior uncertainty is reasonable and well-calibrated.}

\begin{table*}[!htbp]
\centering
\begin{threeparttable}
\caption{
    Comparison of image quality metrics for Fig.~\ref{fig:mclmc mean vs mode and fitting vs mea} with different measurement statistics.
}
\label{tab:mean_vs_mode_at_different_measurement_time}

\begin{tabular}{>{\centering\arraybackslash}m{3.0cm} >{\raggedright\arraybackslash}m{2.5cm} >{\raggedleft\arraybackslash}m{3.2cm} >{\raggedleft\arraybackslash}m{3.2cm} >{\raggedleft\arraybackslash}m{3.2cm}}
\toprule
Method & Metric & $T = 10$~mins & $T = 100$~mins & $T = 1000$~mins \\
\midrule
\multirow{3}{*}{Mean}
& PSNR & 19.44 & 25.42 & 28.82 \\
& SSIM & 0.58 & 0.87 & 0.95 \\
& Total intensity ratio& 1.044 & 1.013 & 1.006 \\
\midrule
\multirow{3}{*}{Marginal mode}
& PSNR & 20.26 & 25.06 & 25.47 \\
& SSIM & 0.75 & 0.87 & 0.88 \\
& Total intensity ratio& 0.626 & 0.797 & 0.844 \\
\midrule
\multirow{3}{*}{Joint mode (MCLMC)}
& PSNR & 18.83 & 22.26 & 24.16 \\
& SSIM & 0.50 & 0.78 & 0.91 \\
& Total intensity ratio& 1.046 & 1.013 & 1.006 \\
\midrule
\multirow{3}{*}{Joint mode (L-BFGS)}
& PSNR & 19.31 & 25.56 & 28.24 \\
& SSIM & 0.57 & 0.87 & 0.94 \\
& Total intensity ratio& 1.050 & 1.013 & 1.006 \\
\bottomrule
\end{tabular}
\begin{tablenotes}[flushleft]
\footnotesize
\item[] \textit{Note:} Total intensity for the ground truth is $8.93\times10^8~\mathrm{emissions~s^{-1}}$ (28.3 mCi assuming Cs-137 source).
\end{tablenotes}

\end{threeparttable}
\end{table*}

\begin{figure}[!htbp]
    \centering
    \includegraphics[width=1\linewidth]{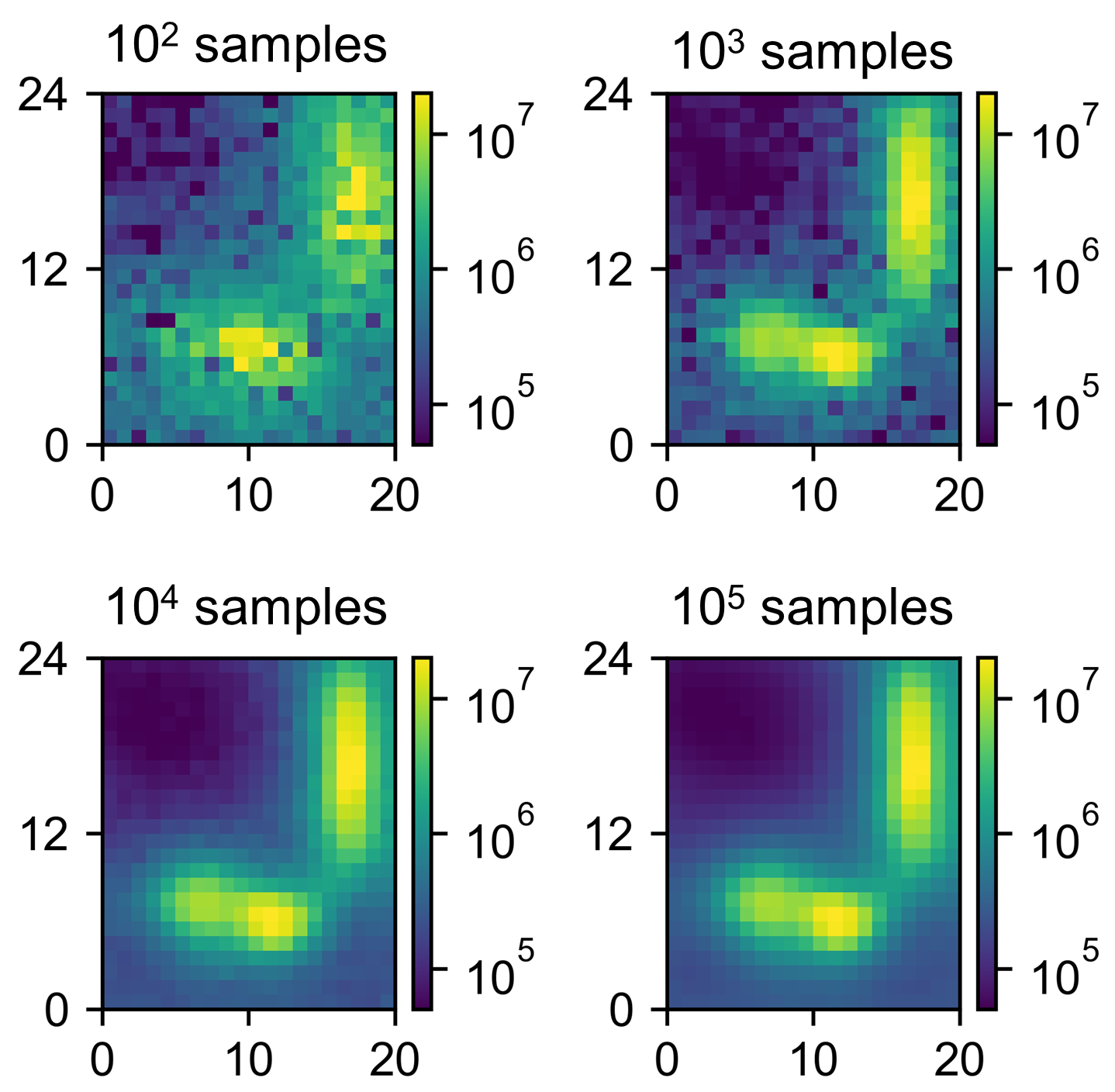}
    \caption{
        Intensity maps using the posterior mean of samples by MCLMC with different numbers of samples, for $T=100$~minutes.
        The PSNR / SSIM / computation time is 17.94 / 0.48 / 8.1~s, 25.09 / 0.84 / 8.3~s, 25.42 / 0.87 / 8.7~s, 25.48 / 0.88 / 14.6~s, for $10^2$, $10^3$, $10^4$, and $10^5$ samples, respectively.
        The computation time includes initialization time.
    }
    \label{fig:mclmc mean vs samples}
\end{figure}

\begin{figure}[!htbp]
    \centering
    \includegraphics[width=1.0\linewidth]{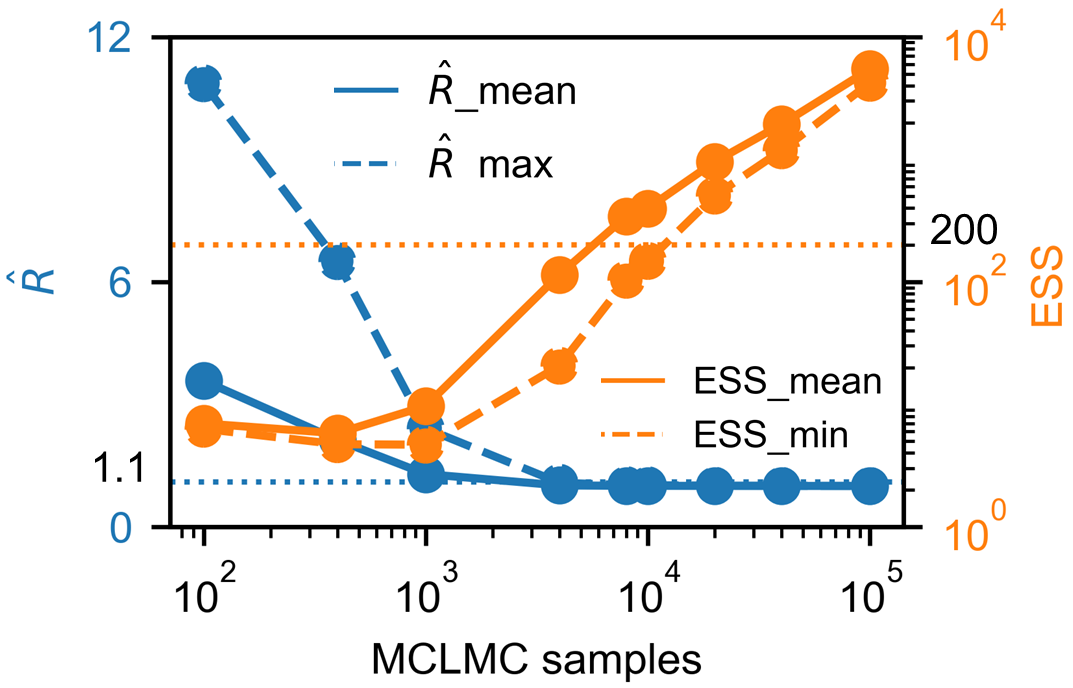}
    \caption{
        $\Rhat$ and ESS for MCLMC with different total number of samples.
        Only $\Rhat$ and ESS for selected number of samples are shown.
        4 chains are used, and the total measurement time is $T=10$~minutes.
    }
    \label{fig:rhat_ess_vs_iterations}
\end{figure}

\begin{figure}[!htbp]
    \centering
    \includegraphics[width=1\linewidth]{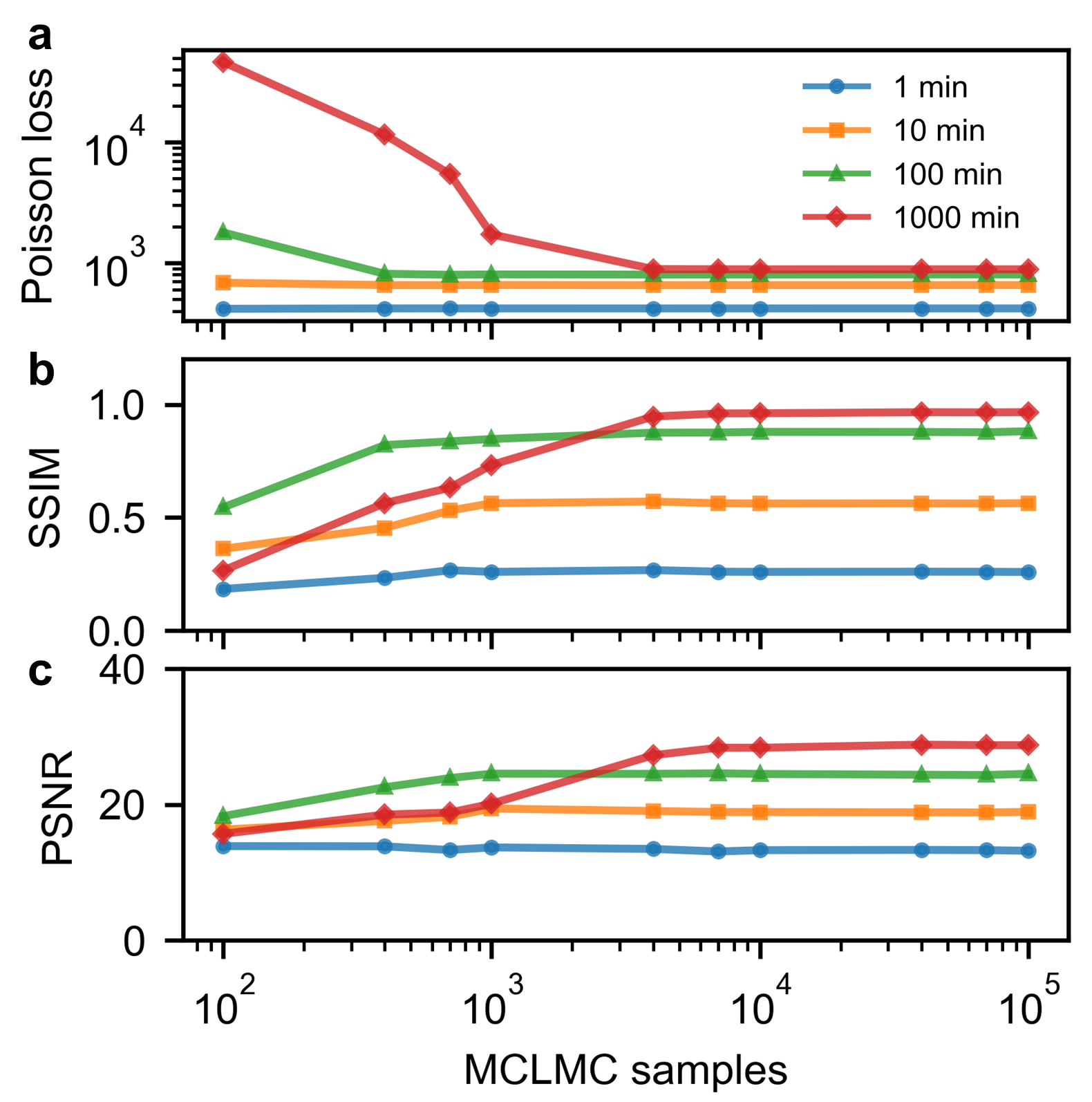}
    \caption{
        (a) Poisson loss calculated using the posterior mean.
        (b),~(c) SSIM and PSNR from comparison of intensity map using the posterior mean against the ground truth with different number of MCLMC samples.
    }
    \label{fig:psnr_ssim_vs_mclmc samples}
\end{figure}

To demonstrate that the samples drawn by MCLMC converge to a stable posterior distribution, qualitatively we can first visually inspect the convergence as a function of samples by examining the intensity maps.
Fig.~\ref{fig:mclmc mean vs samples} shows the intensity map using posterior mean corresponding to $10^2$, $10^3$, $10^4$, and $10^5$ samples.
The intensity map for~$10^2$ samples is noisy and significantly different from that of $10^5$ samples, indicating that~$10^2$ samples is too few to reach convergence in this example.
The mean map of $10^4$ samples is similar to that of $10^5$ samples, which suggests good convergence.
Quantitatively, Fig.~\ref{fig:rhat_ess_vs_iterations} shows the~$\Rhat$ and ESS as function of total number of samples.
Since $\Rhat$ and ESS have the same dimension as the number of pixels~$J$, we plot the mean and maximum value of $\Rhat$ and mean and minimum value of ESS.
As mentioned in Section~\ref{sec:image_reconstruction}, $\Rhat < 1.1$ and $\text{ESS} > 200$ are considered good convergence, indicating that a total number of~${\sim}10^4$ samples is sufficient for this synthetic data scenario with $J = 480$ pixels.
Consequently, we used a total number of~$10^4$ samples for other testing scenarios with this synthetic data unless explicitly noted.
The convergence can also be quantified using the Poisson loss (Eq.~\ref{eq:log_likelihood}) from the forward projection of the reconstructed image, in this case the posterior mean map.
Fig.~\ref{fig:psnr_ssim_vs_mclmc samples}a shows that the Poisson loss decreases as number of samples increases, which means the MCLMC results fit the measured count data well.
Fig.~\ref{fig:psnr_ssim_vs_mclmc samples}b and Fig.~\ref{fig:psnr_ssim_vs_mclmc samples}c show the PSNR and SSIM calculated from comparison of the posterior mean map against the ground truth, which also demonstrates good convergence with ${<}10^4$ samples.
Unlike ML-EM where PSNR and SSIM would decrease due to overfitting as the number of iterations exceeds the optimal choice, MCLMC does not tend to overfit as the number of samples increases, and the PSNR and SSIM remain stable.

We also tested different priors---the truncated Gaussian prior and the Gaussian Process Prior.
For the truncated Gaussian prior, we set the mean and standard deviation of the truncated Gaussian distribution using two different methods.
One method uses the intensity estimate of individual pixels from 10 ML-EM iterations (as shown in Fig.~\ref{fig:mclmc prior and post}) as the mean and standard deviation of corresponding pixels.
The other method sets the mean and standard deviation to be a constant value of $1 \times 10^7$ for all pixels.
{\color{black}For the Gaussian Process prior (GPP), we do not need to manually specify any values.
Instead, there are hyperparameters in the GPP representing the correlation length scale and average prior pixel values. These hyperparameters are determined using an approximate empirical Bayes method based on the Laplace approximation~\cite{lee2024radiation}. The marginal likelihood is optimized with respect to the hyperparameters (known as type-II maximum likelihood). }
{\color{black}In this case, the correlation length is 2 m and the average prior pixel value is $3.89 \times 10^6$ emissions/s after optimization. }

\begin{figure}[!htbp]
    \centering
    \includegraphics[width=1\linewidth]{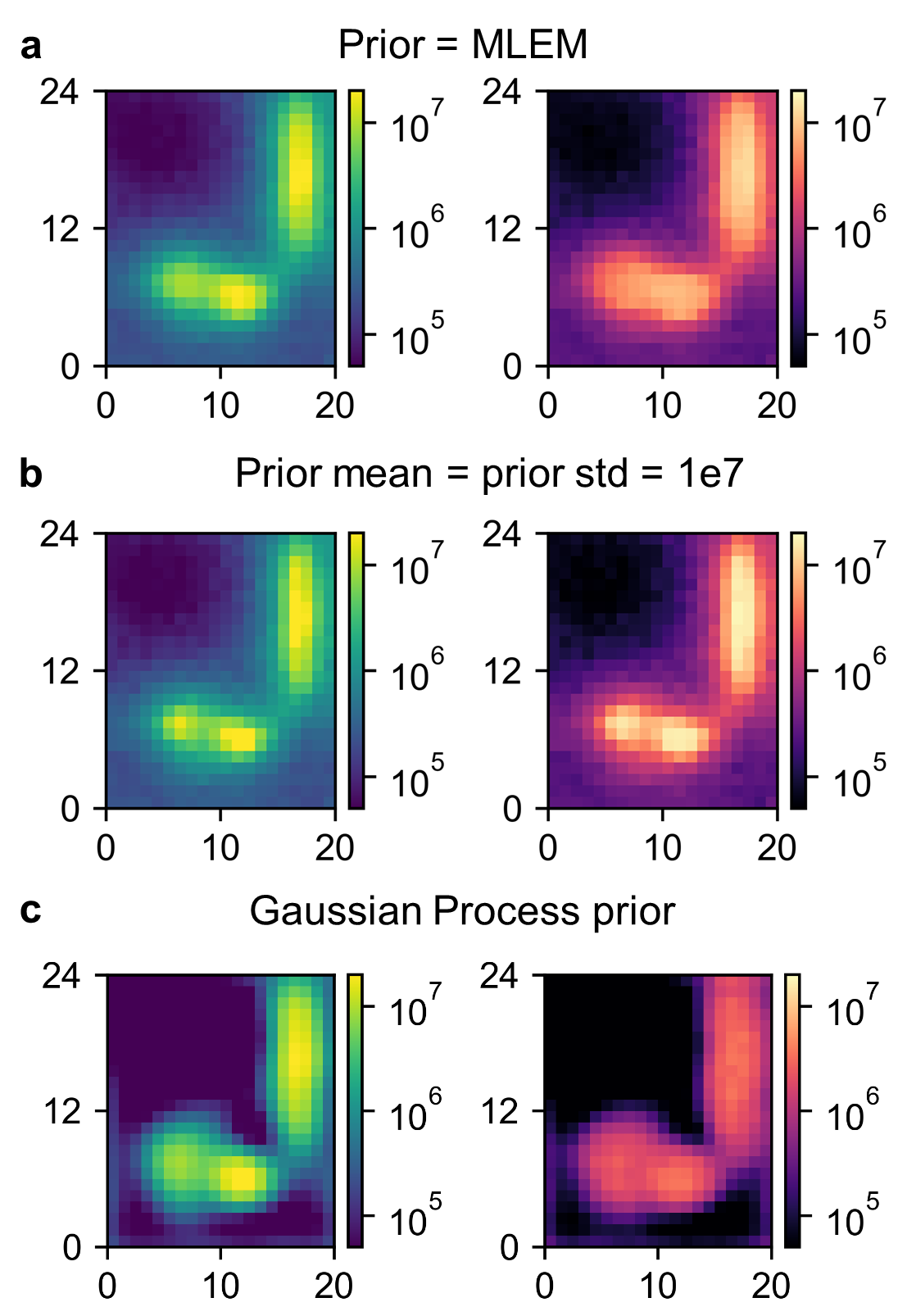}
    \caption{
        (a) Posterior mean and uncertainty map for intensity estimates with 10 ML-EM iterations used as mean and standard deviation of truncated Gaussian prior; PSNR = 25.42; SSIM = 0.87 against ground truth.
        (b) Same, for a flat value of $10^7$ used as mean and standard deviation of truncated Gaussian prior; PSNR = 24.73; SSIM = 0.86.
        (c) Same for the Gaussian process prior; PSNR = 29.05; SSIM = 0.95.
        Total measurement time $T$ = 100 minutes.
        Uncertainty is the 68$\%$ HDI of the marginal distributions.
    }
    \label{fig:recon and UQ MCLMC}
\end{figure}

{\color{black}As shown in Fig.~\ref{fig:recon and UQ MCLMC}, with the same measurement ($T=100$~minutes), the posterior mean intensity map and the uncertainty map are similar when using different truncated Gaussian priors, which suggests that the posterior is insensitive to the number of ML-EM iterations used to set the truncated Gaussian prior given sufficient measurement. }
The posterior mean map obtained using GPP is significantly better than that using the truncated Gaussian prior.
{\color{black}Notably, the posterior mean map from GPP with measurement time $T=100$~minutes has PSNR = 29.05 and SSIM = 0.95, which is better than the PSNR = 28.82 and SSIM = 0.95 for the truncated Gaussian prior with a significantly longer measurement time $T=1000$~minutes (see Table~\ref{tab:mean_vs_mode_at_different_measurement_time}).}
Moreover, the uncertainty map from the GPP is more confined to the source area and on average gives significantly smaller uncertainties than the truncated Gaussian prior.
This is due to the spatial correlations present in the GPP but not the (independent) truncated Gaussian prior---roughly speaking, in the latter case, a single pixel must vary by a large amount to produce the same change in likelihood achieved by perturbing the activities of many correlated pixels.
We note that it is normal that the posteriors corresponding to different priors are not identical, because the prior affects the posterior, and the degree of the prior's effect depends on the measurement statistics.
These results demonstrate that the MCLMC sampler can be used with different types of priors and better intensity and uncertainty estimates can be achieved with a proper type of prior.

\begin{figure}[!htbp]
    \centering
    \includegraphics[width=1\linewidth]{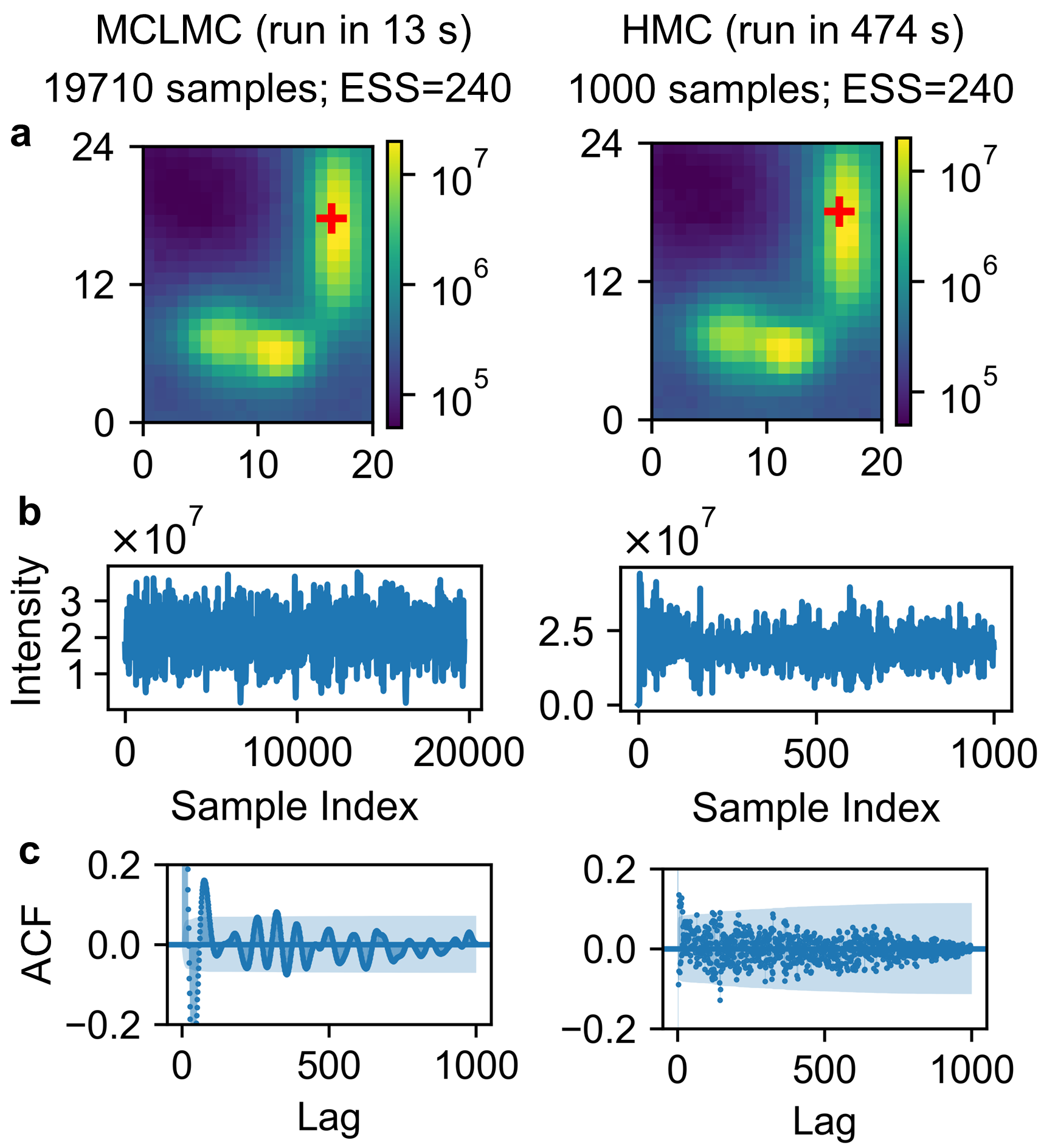}
    \caption{
        Performance comparison of MCLMC vs HMC at similar ESS.
        Top: intensity maps using the posterior mean.
        PSNR = 25.48, SSIM = 0.87 for MCLMC and PSNR = 25.76, SSIM = 0.88 for HMC.
        The pixel in the source region labeled ``$+$'' is selected for the trace and ACF plots.
        Middle: trace plot showing the value of the selected pixel as function of the sample index.
        Bottom: Autocorrelation Function (ACF) plot.
        The shaded region is the error band representing the $95\%$ confidence interval.
        ACF values outside the band suggest significant autocorrelation.
        Total measurement time is $T = 100$~minutes.
    }
    \label{fig:hmc vc mclmc}
\end{figure}

We compared MCLMC against HMC on the same synthetic data example.
As the samples drawn by different MCMC samplers may have different degrees of autocorrelation, it is useful to compare the Effective Sample Size (ESS) per unit computation time rather than imaging performance at a fixed number of samples.
For instance, with a total measurement time $T = 100$~minutes and a fixed number of $10000$ samples, MCLMC has an mean ESS (averaged over all pixels) of $112$ in a computation time of $11.8$~s ($9.5$~ES/s), while HMC has a mean ESS of $2396$ in $5105.1$~s ($0.5$~ES/s).
We also compare performance when the ESS is equated between MCLMC and HMC.
With a total number of 1000 samples, HMC achieves mean ESS of $240$ in a computation time of $474.3$~s ($0.5$~ES/s).
In comparison, MCLMC achieves the same ESS of $240$ at a much larger sample count of $19710$ but in a much shorter computation time of $12.9$~s ($18.6$~ES/s).
Fig.~\ref{fig:hmc vc mclmc} shows the intensity map using the posterior mean from MCLMC and HMC for the same mean ESS of $240$.
The intensity maps have similar image quality metrics of PSNR = 25.48, SSIM = 0.87 for MCLMC and PSNR = 25.76, SSIM = 0.88 for HMC.
The trace plot and ACF plot for a pixel in the source region (same pixel as the pixel labeled ``$+$'' in Fig.~\ref{fig:synthetic_data_ground_truth}) are also shown, which suggests good convergence and similar autocorrelation for MCLMC and HMC.
Further reducing the total number of samples of HMC to $100$ leads to worse convergence, with a mean ESS of $28.4$ in a computation time of $50.3$~s ($0.6$~ES/s).
In conclusion, MCLMC takes significantly shorter time than HMC to achieve a similar ESS.

We found that the measurement time (and therefore the count statistics) could, in some cases, affect the convergence of the MCMC samplers.
With a total measurement time $T = 1000$ minutes, the MCLMC converges well (as shown in Fig.~\ref{fig:mclmc mean vs mode and fitting vs mea}), but HMC failed to converge.
We also tried other MCMC samplers in Blackjax, such as NUTS, but these were unsuccessful for our case across a reasonable range of parameter tunings.
The MCLMC significantly outperforms other MCMC samplers and it is promising for near-real-time uncertainty quantification.

It is important to run the MCLMC in a time as short as possible to allow near-real-time uncertainty quantification, so we investigated running multiple chains in parallel to more quickly reach the number of samples required for good convergence.
Specifically, we tested two acceleration options available via Blackjax, namely CPU multi-threading and GPU acceleration.
We note that the CPU multi-threading is not fully parallel across the CPU cores, as the threads share the same process, but it could save some time compared to running the chains fully sequentially.
In comparison, the GPU offers true parallel processing of different chains.
The CPU configuration used here is an AMD Ryzen Threadripper 7960X processor ($24$ cores, $48$ threads, x86\_64 architecture) with 128 GB RAM, and the GPU is an NVIDIA GeForce RTX 4090 ($24$~GB memory with CUDA version 12.8).

Table~\ref{tab:sampling time_synthetic data} lists two times, $t_a$, $t_b$, for each setting, where~$t_a$ includes the initialization or warm-up time for the first run, and~$t_b$ represents the second run where no initialization is needed.
Here, we run same setting multiple times for bookkeeping, though this is may not be necessary in practice.
Each setting was run 5 times and the average run time $\pm$ standard deviation is shown.
With same total 10000 samples, CPU multi-threading takes more time with more chains due to the overhead of running each chain.
In comparison, when number of chains increased, the first GPU run takes slightly more time for more chains (compare $t_a =9.7 \pm 0.1$~s for one chain against $t_a = 10.6 \pm 0.3$~s for 10 chains), but the second run takes similar time (compare $t_b=3.7 \pm 0.2$~s for 1 chain against $t_b = 3.4 \pm 0.1$~s for 10 chains).

\begin{table}[!htbp]

\caption{Computation time comparison: CPU multi-threading (MT) vs GPU ($J=480$ pixels for the synthetic data in Fig.~\ref{fig:synthetic_data_ground_truth})}
\label{tab:sampling time_synthetic data}
    \centering
    \renewcommand{\arraystretch}{0.9} % optional: slightly tighter rows
    \begin{tabular}{clll}
        \hline
        Total samples & \# chains & \makecell{Run time---\\CPU MT} & \makecell{Run time---\\GPU} \\
        \hline
  10000 & 1 &
    \makecell{$t_a=8.2 \pm 0.4$~s \\$t_b=3.1 \pm 0.2$~s}&
    \makecell{$t_a=9.7 \pm 0.1$~s \\$t_b=3.7 \pm 0.2$~s}\\
  \hline
  10000 & 10 &
    \makecell{$t_a=13.1 \pm 0.5$~s \\$t_b=7.9 \pm 0.3$~s}&
    \makecell{$t_a=10.6 \pm 0.3$~s \\$t_b=3.4 \pm 0.1$~s}\\
  \hline
    \end{tabular}
\end{table}

\subsection{Measured radiological mapping data}\label{sec:results_measured}

\begin{figure}[!htbp]
    \centering
    \includegraphics[width=1\linewidth]{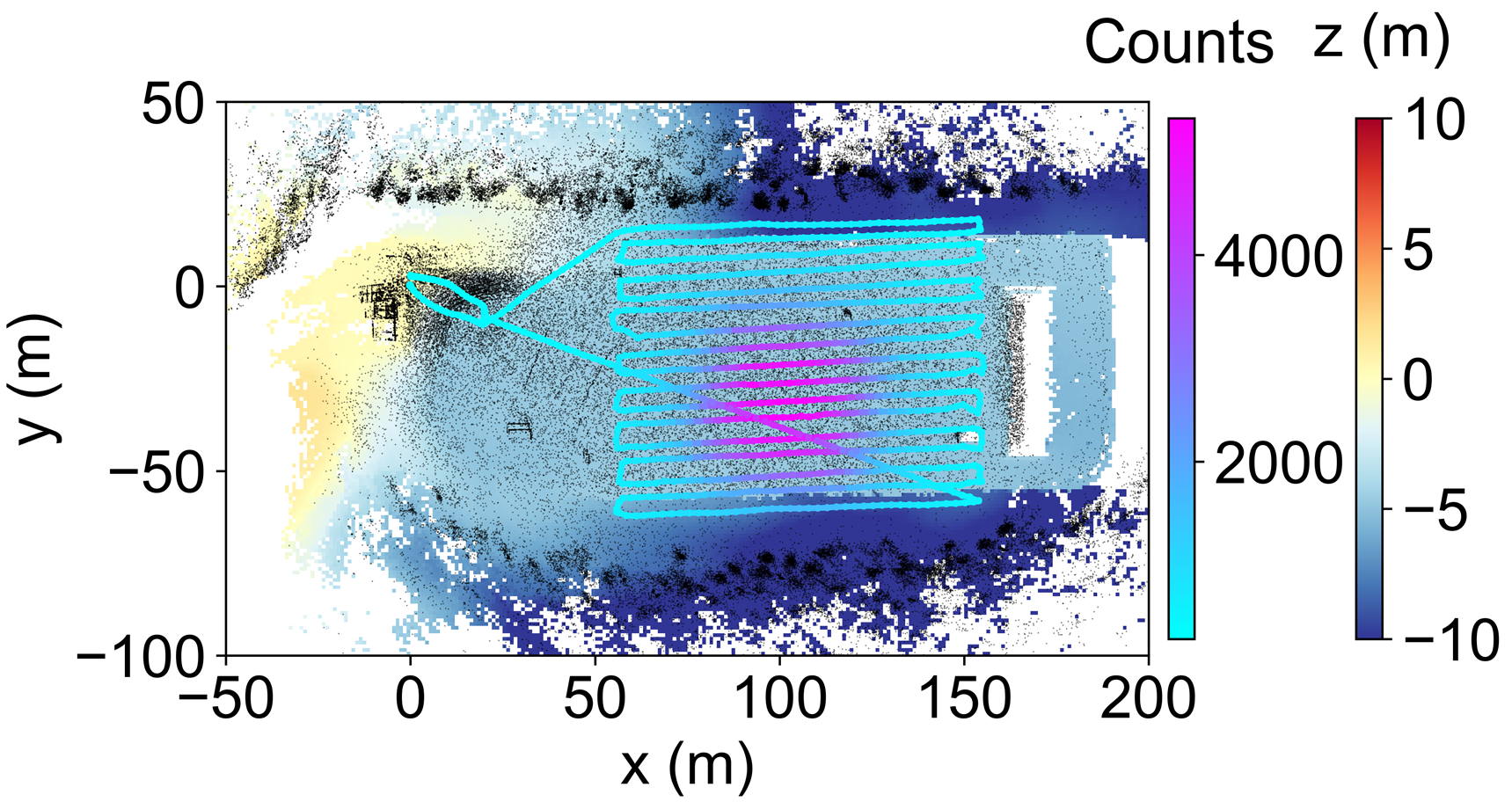}
    \caption{
        Real radiological mapping data.
        The ground height estimate~$z$ has a $1$~m pixel size.
        The trajectory covers the area where the distributed sources are, and is colorized by the gross counts at discretized measurement positions.
        The black dots show the top-down view of the lidar point cloud; trees and buildings are visible upon inspection.
    }
    \label{fig:ground_surface}
\end{figure}

We demonstrate the image reconstruction and UQ performance of MCLMC for measured data from a 2023 aerial mapping campaign at the Johns Hopkins University Applied Physics Laboratory (JHU APL).
A surrogate distributed source~\cite{vavrek2024surrogateI} consisting of a $4$-m-pitch $10 \times 10$ grid of Cs-137 point sources (total activity ${\sim}91$~mCi) was deployed on flat terrain.
The SDF-capable detector system NG-LAMP~\cite{Pavlovsky2019} was flown at $2.5$~m/s at $6$~m above ground level in a $5$-m-spaced raster pattern over the source, with a total measurement time of~${\sim}15$~minutes.
An energy region of interest of $[612, 712]$~keV was used to isolate $662$~keV photopeak events, and data streams from three active detector modules were co-added.
The ground surface was estimated from the lidar point cloud using the Cloth Simulation Filter method~\cite{zhang2016easy, csf_github} at a pixel size of $1$~m, giving $J \simeq 30\,000$~pixels for entire ground surface.
Fig. \ref{fig:ground_surface} shows the ground height map with a pixel size of $1$~m, the detector trajectory at time discretization of $t_i = 1$~s, and the measured counts at discretized measurement positions.
The reconstruction and UQ can be performed for the entire ground surface map or for a smaller area closer to the raster pattern trajectory.
In areas that are far from the measurement trajectory, the posterior is expected to be similar to the prior due to the lack of measurements with which to update the prior.
Therefore we focus most of our analysis of image reconstruction and UQ on the (nearly planar) area within the bounds of the raster trajectory where $J \simeq 10\,000$.
For the later computation time evaluation, however, we still attempt to use the full $J \simeq 30\,000$~pixel problem as a more computationally demanding benchmark, though GPP priors proved challenging due to RAM limitations on our computer.

\begin{figure*}[!htbp]
    \centering
    \includegraphics[width=1\linewidth]{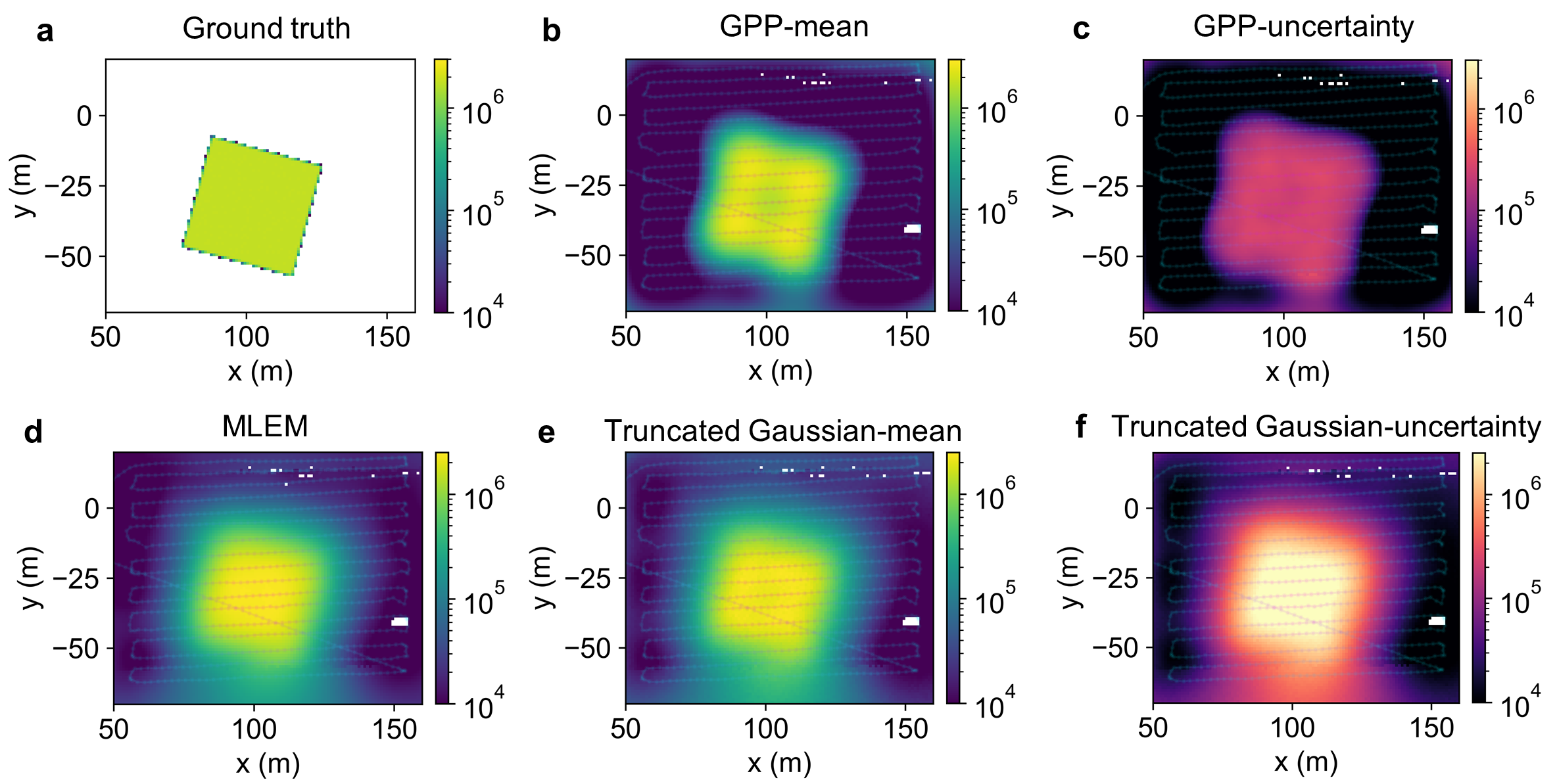}
    \caption{
        MCLMC reconstruction results for the measurement in Fig.\ref{fig:ground_surface}.
        (a) Expected activity map.
        (b),~(c) Mean and uncertainty map with GPP used.
        (d) Intensity reconstruction by ML-EM (20 iterations).
        (e),~(f) Mean and uncertainty map with truncated Gaussian prior used, and mean and standard deviation of the truncated Gaussian prior is set to be equal to the intensity reconstruction by ML-EM 20 iterations in (d).
    }
    \label{fig:square_source_mlem_mean_UQ}
\end{figure*}

By design~\cite{vavrek2024surrogateI}, the ground truth intensity distribution of the $4$-m-pitch Cs-137 point source grid can be well-approximated with the continuous or ``interpolated'' $36\, \text{m} \times 36\, \text{m}$ square source shown in Fig.~\ref{fig:square_source_mlem_mean_UQ}a.
The total activity of the expected activity map is set to 91 mCi, corresponding to $2.87 \times 10^{9}$ emissions/s of the $662$~keV gamma-ray using the branching ratio of $85\%$ for the $662$~keV gamma-rays of Cs-137 \cite{leblond2024ddep}.

We tested MCLMC-based reconstruction using two different priors, the truncated Gaussian prior and the Gaussian Process prior (GPP).
For the truncated Gaussian prior, we used the intensity estimates from 20 ML-EM iterations as the mean and standard deviation, as shown in Fig.~\ref{fig:square_source_mlem_mean_UQ}d.
The choice of $20$ ML-EM iterations instead of $10$ was made to improve the quality of the ML-EM result.
{\color{black}For GPP, the hyperparameters determined by the approximate empirical Bayes method are as follows: correlation length = 12 m and mean prior pixel value = $5.7 \times 10^5$ emissions/s.}
Comparing the reconstructed intensity map using the posterior mean against the expected activity map, the GPP yields PSNR=16.38 and SSIM=0.66, with a total intensity of $3.26 \times 10^{9}$ emissions/s, i.e., $1.14\times$ the estimated total activity.
The truncated Gaussian prior yields PSNR=15.38 and SSIM=0.35, with a total intensity of $3.50 \times 10^{9}$ emissions/s, i.e., $1.22\times$ the estimated total activity (compare Fig.~\ref{fig:square_source_mlem_mean_UQ}b vs.\ Fig.~\ref{fig:square_source_mlem_mean_UQ}e).
The uncertainty map obtained from the GPP is better spatially-confined to the source area, and the uncertainty values themselves are significantly smaller than the uncertainties from the truncated Gaussian prior, again reflecting the spatial correlations inherent in the GPP.

\begin{figure}[!htbp]
    \centering
    \includegraphics[width=1\linewidth]{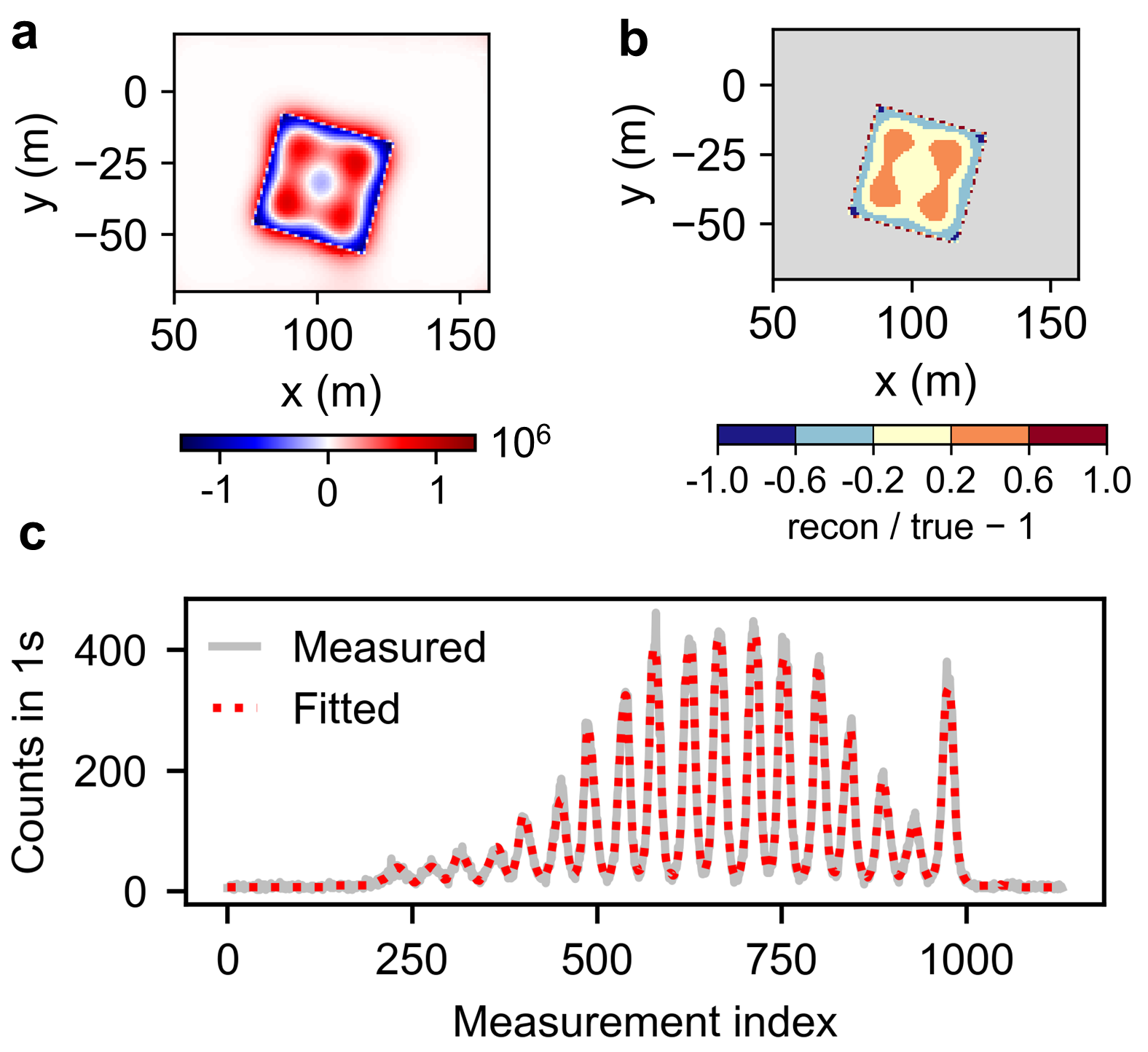}
    \caption{
        (a) Absolute and (b) relative difference map between the reconstructed intensity with GPP and the expected activity map.
        (c) Measurement vs.\ fitted mean counts calculated from Eq.~\ref{eq:lambda} using the posterior mean intensity values with GPP.
    }
    \label{fig:mclmc_counts_and_intensities}
\end{figure}

When the truncated Gaussian prior is used, the posterior is very similar to the prior (compare Fig.~\ref{fig:square_source_mlem_mean_UQ}d and Fig.~\ref{fig:square_source_mlem_mean_UQ}e), which implies that the measured data is not sufficient to significantly update the prior when the truncated Gaussian prior is used.
In contrast, the GPP produced better performance for the same measurement data.
Absolute and relative difference maps for the GPP MCLMC reconstruction compared to the expected activity map are shown in Fig.~\ref{fig:mclmc_counts_and_intensities}.
The reconstructed intensities in a ring around the center of the source and just beyond the the source extents are larger than the expected activity map, while the source center and particularly the periphery near the exterior of the source are smaller.
This is a result of the smooth transition of the reconstructed intensity.
Fig.~\ref{fig:mclmc_counts_and_intensities}c shows that the fitted mean counts, calculated from Eq.~\ref{eq:lambda} using the posterior mean intensity values with GPP, agree well with the measured counts, which indicates that the reconstruction results by MCLMC with GPP adequately explain the measured data.

Using the same computer hardware as in Sec.~\ref{sec:results_recon}, we also compared the run time when truncated Gaussian prior and GPP are used.
With the truncated Gaussian prior, we ran radiation image reconstructions for the entire ground surface with $J \simeq 30\,000$ pixels (Table~\ref{tab:sampling time_real data}).
When a total of 10000 samples is run in 1~chain, GPU takes a shorter time ($16.6 \pm 0.8$~s) than CPU multi-threading ($18.8 \pm 0.4$~s).
With the same total of 10000 samples run in 10 chains, the computation time for CPU multi-threading increases.
Because there is initialization needed to set up a chain before drawing the samples, running more chains overall needs more computation than simply adding more samples to a chain.
In comparison, when the chains are run in parallel on the GPU, the overall computation time decreases when the total number of samples is held constant but split across more chains.
With parallel processing on the GPU, for a constant 1000 samples per chain, a total of 1000 samples run in 1~chain takes similar time as that of 10000 samples in 10 chains.
For CPU multi-threading, the time for 10000 samples in 10 chains is longer than that of 1000 samples in 1~chain.

With the GPP, we could only run reconstructions for a smaller area within the bounds of the trajectory ($J \simeq 10\,000$ pixels) since running more pixels leads to out-of-memory errors.
When a total of 10000 samples is run in 1 chain, CPU multi-threading takes $33.9 \pm 0.1$~s / $26.0 \pm 0.1$~s, while the GPU takes $32.0 \pm 0.1$~s / $26.5 \pm 0.1$~s for the first/second run.

It is also worthwhile to compare the effect of increased dimension on the computation time.
Despite the large increase from the synthetic data $J = 480$ in Table~\ref{tab:sampling time_synthetic data} to the real data $J \simeq 30\,000$ in Table~\ref{tab:sampling time_real data}, the computation time does not increase dramatically using the same computer hardware.
Remarkably, with $10\,000$ samples in 10 chains run on GPU, the computation time for $J \simeq 30\,000$  is similar to that of $J = 480$.
These results show that MCLMC is well-suited for high dimensional image reconstruction and UQ, and that running MCLMC in parallel on the GPU is an effective method for reducing the computation time, favoring the GPU for near-real-time operation.

\begin{table}[!htbp]
\caption{Computation time comparison for the truncated Gaussian prior: CPU multi-threading (MT) vs GPU ($J \simeq 30\,000$~pixels for the measurement in Fig.~\ref{fig:ground_surface})}
\label{tab:sampling time_real data}
    \centering
\renewcommand\cellalign{tl}
\begin{tabular}{cllll>{\raggedright\arraybackslash}m{3.8cm}>{\raggedright\arraybackslash}m{3.8cm}}
\toprule
Total samples & \# chains  & Run time---CPU MT & Run time---GPU \\
\midrule
10000 & 1 & 
\makecell[tl]{$t_a$ = 18.8 $\pm$ 0.4 s\\$t_b$ = 12.9 $\pm$ 0.2 s} &
\makecell[tl]{$t_a$ = 16.6 $\pm$ 0.8 s\\$t_b$ = 9.2 $\pm$ 0.2 s} \\\hline

10000 & 10 & 
\makecell[tl]{$t_a$ = 24.7 $\pm$ 1.3 s\\$t_b$ = 16.6 $\pm$ 0.7 s} &
\makecell[tl]{$t_a$ = 11.2 $\pm$ 1.0 s\\$t_b$ = 3.7 $\pm$ 0.1 s} \\\hline

1000 & 1 & 
\makecell[tl]{$t_a$ = 11.8 $\pm$ 0.3 s\\$t_b$ = 6.3 $\pm$ 0.1 s} &
\makecell[tl]{$t_a$ = 10.5 $\pm$ 0.5 s\\$t_b$ = 3.7 $\pm$ 0.2 s} \\\hline

1000 & 10 & 
\makecell[tl]{$t_a$ = 21.8 $\pm$ 1.4 s\\$t_b$ = 14.3 $\pm$ 0.5 s} &
\makecell[tl]{$t_a$ = 10.7 $\pm$ 0.9 s\\$t_b$ = 3.3 $\pm$ 0.1 s} \\
\bottomrule
\end{tabular}

\end{table}

\section{Discussion}
The results above demonstrate that the MCLMC algorithm is a promising method for improving distributed radiological source imaging and providing computationally-efficient UQ estimates.
Here we discuss some limitations of the method and opportunities to explore or employ it further in future work.

One practical limitation of MCMC-based methods is that typically the entire system matrix~$\boldsymbol{V}$ must be constructed and stored in memory (either RAM or GPU VRAM).
For large mapping areas and high spatial fidelity in both the scene model and the detector position, respectively, the number of pixels/voxels~$J$ and number of measurement timestamps~$I$ can be large enough that the $I \times J$ system matrix exceeds the capacity of typical consumer hardware, especially if distinct positions are maintained for individual detector modules in multi-module systems ($I \to I \cdot N_\text{modules}$).
It may however be acceptable to run the UQ at lower spatial fidelity, or only within smaller regions of the scene covered by the measurement trajectory such as in Section~\ref{sec:results_measured}.

Additionally, the present Blackjax GPU implementation relies on CUDA.
Systems without NVIDIA GPUs will likely need to fall back on the slower multi-threaded CPU implementation, which may be challenging for near-real-time operation.

In future work, a dedicated head-to-head comparison of various potential distributed source UQ methods and priors---including the Laplace and pCN-MCMC methods~\cite{lee2024radiation}, and, if possible, IBU~\cite{bourbeau2018pyunfold}---would be valuable for better understanding the implications of the different absolute scales of uncertainty maps observed.
{\color{black}We also focused on distributed sources that are characterized by spatial smoothness, which is typical in wide-area contamination mapping. It would also be interesting to test sources with sharp edges using a prior that incorporates structural information.}

{\color{black}Furthermore, the recent advances in MCMC samplers, including the MCLMC algorithm, open up the possibility for fully Bayesian approaches in high-dimensional imaging problems. Fully Bayesian treatment places prior distributions on hyperparameters governing the prior over the image, and the image and the hyperparameters are jointly inferred. Such an approach appropriately propagates uncertainties in hyperparameters, thereby yielding better uncertainty estimation, particularly in weakly constrained regions.
Though such an approach has not been explored for radiation mapping problems due to its computational demand, recent MCMC techniques---including MCLMC---can enable efficient sampling from high-dimensional joint posteriors over both image and hyperparameters, making it a promising direction for future work.}

It would also be valuable to demonstrate the use of the MCLMC UQ results to drive decision-making algorithms such as path planning through or around contamination.
Near-real-time uncertainty maps could help define data sufficiency criteria such as measuring until every pixel uncertainty falls below some absolute threshold, or path planning strategies such as moving towards regions with the largest activity uncertainty, either autonomously or via feedback to human operators.
Performing MCLMC UQ live on SDF-capable detector systems during distributed source mapping experiments would therefore be a logical next step to operationalizing the algorithm.

\section{Conclusions}

We applied a newly-developed MCLMC sampler to improve the accuracy and stability of distributed radiological source imaging, as well as provide UQ estimates.
In both synthetic and measured datasets, MCLMC was found to be extremely computationally efficient compared to prior state-of-the-art MCMC samplers such as Hamiltonian Monte Carlo, finding images and Bayesian credible intervals in runtimes on the order of ${\sim}10$~s or less.
The MCLMC-reconstructed image tends to converge towards the true image given sufficient measurement data and a reasonable prior, without requiring a user-defined stopping criterion to avoid overfitting as in ML-EM.
More informative priors, such as the Gaussian Process Prior (GPP), can improve the reconstruction when data is limited.
MCLMC therefore enables near-real-time accuracy improvements and uncertainty quantification for radiation mapping, which in turn can provide rapid feedback for radiological emergency response and nuclear security applications.

\section*{Acknowledgments}
The authors thank Simon Mak (Duke University) and Mark Bandstra (LBNL) for useful discussions.
This manuscript has been authored by an author at Lawrence Berkeley National Laboratory under Contract No.\ DE-AC02-05CH11231 with the U.S.\ Department of Energy.
The U.S.\ Government retains, and the publisher, by accepting the article for publication, acknowledges, that the U.S.\ Government retains a non-exclusive, paid-up, irrevocable, world-wide license to publish or reproduce the published form of this manuscript, or allow others to do so, for U.S.\ Government purposes.
The views and opinions of authors expressed herein do not necessarily state or reflect those of the United States Government or any agency thereof or the Regents of the University of California.

\bibliographystyle{IEEEtran}
\bibliography{references}

@article{leblond2024ddep,
  title={{DDEP} re-evaluation of the radioactive decay scheme of {137Cs}},
  author={Leblond, Sylvain},
  journal={Applied Radiation and Isotopes},
  volume={206},
  pages={111191},
  year={2024},
  publisher={Elsevier}
}

@article{rue2009approximate,
  title={Approximate {Bayesian} inference for latent {Gaussian} models by using integrated nested {Laplace} approximations},
  author={Rue, H{\aa}vard and Martino, Sara and Chopin, Nicolas},
  journal={Journal of the Royal Statistical Society Series B: Statistical Methodology},
  volume={71},
  number={2},
  pages={319--392},
  year={2009},
  publisher={Oxford University Press}
}

@book{williams2006gaussian,
  title={Gaussian processes for machine learning},
  author={Williams, Christopher KI and Rasmussen, Carl Edward},
  volume={2},
  number={3},
  year={2006},
  publisher={MIT press Cambridge, MA}
}

@article{roy2020convergence,
  title={Convergence diagnostics for {Markov chain Monte Carlo}},
  author={Roy, Vivekananda},
  journal={Annual Review of Statistics and Its Application},
  volume={7},
  number={1},
  pages={387--412},
  year={2020},
  publisher={Annual Reviews}
}

@inproceedings{hore2010image,
  title={Image quality metrics: {PSNR} vs.\ {SSIM}},
  author={Hore, Alain and Ziou, Djemel},
  booktitle={2010 20th international conference on pattern recognition},
  pages={2366--2369},
  year={2010},
  organization={IEEE}
}

@article{vetter2018gamma,
  title={Gamma-Ray imaging for nuclear security and safety: Towards {3-D} gamma-ray vision},
  author={Vetter, Kai and Barnowksi, Ross and Haefner, Andrew and Joshi, Tenzing HY and Pavlovsky, Ryan and Quiter, Brian J},
  journal={Nuclear Instruments and Methods in Physics Research Section A: Accelerators, Spectrometers, Detectors and Associated Equipment},
  volume={878},
  pages={159--168},
  year={2018},
  publisher={Elsevier}
}

@article{vetter2019advances,
  title={Advances in nuclear radiation sensing: Enabling {3-D} gamma-ray vision},
  author={Vetter, Kai and Barnowski, Ross and Cates, Joshua W and Haefner, Andrew and Joshi, Tenzing HY and Pavlovsky, Ryan and Quiter, Brian J},
  journal={Sensors},
  volume={19},
  number={11},
  pages={2541},
  year={2019},
  publisher={MDPI}
}

@article{hellfeld2019gamma,
  title={Gamma-ray point-source localization and sparse image reconstruction using {Poisson} likelihood},
  author={Hellfeld, Daniel and Joshi, Tenzing HY and Bandstra, Mark S and Cooper, Reynold J and Quiter, Brian J and Vetter, Kai},
  journal={IEEE Transactions on Nuclear Science},
  volume={66},
  number={9},
  pages={2088--2099},
  year={2019},
  publisher={IEEE}
}

@article{dempster1977maximum,
  title={Maximum likelihood from incomplete data via the {EM} algorithm},
  author={Dempster, Arthur P and Laird, Nan M and Rubin, Donald B},
  journal={Journal of the Royal Statistical Society: Series B (Methodological)},
  volume={39},
  number={1},
  pages={1--22},
  year={1977},
  publisher={Wiley Online Library}
}

@article{vavrek2024surrogateI,
  title={Surrogate Distributed Radiological Sources—Part~{I}: Point-Source Array Design Methods},
  author={Vavrek, Jayson R and Bandstra, Mark S and Hellfeld, Daniel and Quiter, Brian J and Joshi, Tenzing HY},
  journal={IEEE Transactions on Nuclear Science},
  volume={71},
  number={2},
  pages={213--223},
  year={2024},
  publisher={IEEE}
}

@article{vavrek2024surrogateIII,
  title={Surrogate distributed radiological sources~{III}: quantitative distributed source reconstructions},
  author={Vavrek, Jayson R and Lee, Jaewon and Salathe, Marco and Bandstra, Mark S and Hellfeld, Daniel and Quiter, Brian J and Joshi, Tenzing HY},
  journal={arXiv preprint arXiv:2412.02926},
  year={2024}
}

@InProceedings{pmlr-v253-robnik24a,
  title = 	 {Fluctuation without dissipation: {Microcanonical Langevin Monte Carlo}},
  author =       {Robnik, Jakob and Seljak, Uros},
  booktitle = 	 {Proceedings of the 6th Symposium on Advances in Approximate Bayesian Inference},
  pages = 	 {111--126},
  year = 	 {2024},
  editor = 	 {Antorán, Javier and Naesseth, Christian A.},
  volume = 	 {253},
  series = 	 {Proceedings of Machine Learning Research},
  month = 	 {21 Jul},
  publisher =    {PMLR},
  url = 	 {https://proceedings.mlr.press/v253/robnik24a.html}
}

@misc{blackjax_mclmc,
    author={{The Sampling Book Project}},
    title={{Microcanonical Langevin Monte Carlo}},
    note={Retrieved April 22, 2025 from \url{https://blackjax-devs.github.io/sampling-book/algorithms/mclmc.html}}
}

@article{zhang2016easy,
  title={An easy-to-use airborne LiDAR data filtering method based on cloth simulation},
  author={Zhang, Wuming and Qi, Jianbo and Wan, Peng and Wang, Hongtao and Xie, Donghui and Wang, Xiaoyan and Yan, Guangjian},
  journal={Remote sensing},
  volume={8},
  number={6},
  pages={501},
  year={2016},
  publisher={MDPI}
}

@misc{csf_github,
    title={{CSF}},
    author={Qi, Jianbo},
    note={Retrieved February 2, 2025 from \url{https://github.com/jianboqi/CSF}},
    year={2024},
}

@article{hellfeld2021free,
  title={Free-moving quantitative gamma-ray imaging},
  author={Hellfeld, D and Bandstra, MS and Vavrek, JR and Gunter, DL and Curtis, JC and Salathe, M and Pavlovsky, R and Negut, V and Barton, PJ and Cates, JW and others},
  journal={Scientific reports},
  volume={11},
  number={1},
  pages={1--14},
  year={2021},
  publisher={Nature Publishing Group}
}

@article{bourbeau2018pyunfold,
  title={{PyUnfold: A Python package for iterative unfolding}},
  author={Bourbeau, James and Hampel-Arias, Zigfried},
  journal={arXiv preprint arXiv:1806.03350},
  year={2018}
}

@conference{vavrek2021mcmc,
    author = {Vavrek, JR and Bandstra, MS and Hellfeld, D and Quiter, BJ and Meehan, K and Barton, PJ and Cates, JW and Moran, A and Negut, V and Pavlovsky, R and Joshi, THY},
    booktitle = {IEEE SORMA West 2021},
    title = {{Markov} Chain {Monte Carlo} Uncertainty Quantification for Shielded Point Source Localization},
    year = {2021}
}

@article{lee2024radiation,
  title={Radiation image reconstruction and uncertainty quantification using a {Gaussian} process prior},
  author={Lee, Jaewon and Joshi, Tenzing H and Bandstra, Mark S and Gunter, Donald L and Quiter, Brian J and Cooper, Reynold J and Vetter, Kai},
  journal={Scientific Reports},
  volume={14},
  number={1},
  pages={22958},
  year={2024},
  publisher={Nature Publishing Group UK London}
}

@article{zhou2020bayesian,
  title={Bayesian inference and uncertainty quantification for medical image reconstruction with {Poisson} data},
  author={Zhou, Qingping and Yu, Tengchao and Zhang, Xiaoqun and Li, Jinglai},
  journal={SIAM Journal on Imaging Sciences},
  volume={13},
  number={1},
  pages={29--52},
  year={2020},
  publisher={SIAM}
}

@article{bardsley2012mcmc,
  title={{MCMC}-based image reconstruction with uncertainty quantification},
  author={Bardsley, Johnathan M},
  journal={SIAM Journal on Scientific Computing},
  volume={34},
  number={3},
  pages={A1316--A1332},
  year={2012},
  publisher={SIAM}
}

@article{dione2024utilizing,
  title={Utilizing {Bayesian} Modeling and {MCMC} for Accurate Characterization of Naturally Occurring Radionuclides Reference Background Levels in Mining Areas},
  author={Dione, Djicknack and Faye, Papa Macoumba and Ndiaye, Nogaye and Sy, Moussa Hamady and Ndiaye, Oumar and Traor{\'e}, Alassane and Ndao, Ababacar Sadikhe},
  journal={World Journal of Nuclear Science and Technology},
  volume={14},
  number={4},
  pages={179--187},
  year={2024},
  publisher={Scientific Research Publishing}
}

@inproceedings{higdon1999bayesian,
  title={Bayesian inference and {Markov chain Monte Carlo} in imaging},
  author={Higdon, David M and Bowsher, James E},
  booktitle={Medical Imaging 1999: Image Processing},
  volume={3661},
  pages={2--11},
  year={1999},
  organization={SPIE}
}

@article{liu1989limited,
  title={On the limited memory {BFGS} method for large scale optimization},
  author={Liu, Dong C and Nocedal, Jorge},
  journal={Mathematical programming},
  volume={45},
  number={1},
  pages={503--528},
  year={1989},
  publisher={Springer}
}

@article{calvetti2018inverse,
  title={Inverse problems: From regularization to {Bayesian} inference},
  author={Calvetti, Daniela and Somersalo, Erkki},
  journal={Wiley Interdisciplinary Reviews: Computational Statistics},
  volume={10},
  number={3},
  pages={e1427},
  year={2018},
  publisher={Wiley Online Library}
}

@article{wang2004image,
  title={Image quality assessment: from error visibility to structural similarity},
  author={Wang, Zhou and Bovik, Alan C and Sheikh, Hamid R and Simoncelli, Eero P},
  journal={IEEE transactions on image processing},
  volume={13},
  number={4},
  pages={600--612},
  year={2004},
  publisher={IEEE}
}

@article{Pavlovsky2019,
  title={{3D} Gamma-ray and Neutron Mapping in Real-Time with the {Localization and Mapping Platform} from Unmanned Aerial Systems and Man-Portable Configurations},
  author={Pavlovsky, R and Cates, JW and Vanderlip, WJ and Joshi, THY and Haefner, A and Suzuki, E and Barnowski, R and Negut, V and Moran, A and Vetter, K and others},
  journal={arXiv:1908.06114},
  year={2019}
}

\end{document}